\documentclass[a4paper,11pt,reqno]{amsart}
\usepackage[left=26mm,right=26mm,top=30mm,bottom=30mm]{geometry}

\usepackage{amssymb}
\usepackage[shortlabels]{enumitem}
\usepackage{bm}
\usepackage{mathtools}
\usepackage{mathrsfs}
\usepackage{array} 
\usepackage{xcolor} 
\usepackage{xparse}
\usepackage[sort&compress,numbers]{natbib}
\usepackage{caption}
\usepackage{subcaption}
\usepackage{booktabs}
\usepackage[section]{placeins}
\usepackage{hyperref}

\theoremstyle{plain}

\numberwithin{equation}{section}

\makeatletter
\newcommand\makebig[2]{%
\@xp\newcommand\@xp*\csname#1\endcsname{\bBigg@{#2}}%
\@xp\newcommand\@xp*\csname#1l\endcsname{\@xp\mathopen\csname#1\endcsname}%
\@xp\newcommand\@xp*\csname#1r\endcsname{\@xp\mathclose\csname#1\endcsname}%
}
\makeatother

\makebig{biggg} {3.0}
\makebig{Biggg} {3.5}
\makebig{bigggg}{4.0}
\makebig{Bigggg}{4.5}
\makebig{biggggg}{5}
\makebig{Biggggg}{5.5}
\makebig{bigggggg}{6}
\makebig{Bigggggg}{6.5}

\makeatletter
\newcommand{\doublehat}[1]{%
\begingroup%
\let\macc@kerna\z@%
\let\macc@kernb\z@%
\let\macc@nucleus\@empty%
\hat{\mathchoice%
{\raisebox{.2ex}{\vphantom{\ensuremath{\displaystyle #1}}}}%
{\raisebox{.2ex}{\vphantom{\ensuremath{\textstyle #1}}}}%
{\raisebox{.16ex}{\vphantom{\ensuremath{\scriptstyle #1}}}}%
{\raisebox{.14ex}{\vphantom{\ensuremath{\scriptscriptstyle #1}}}}%
\smash{\hat{#1}}}%
\endgroup%
}
\makeatother

\NewCommandCopy{\ordinaryexists}{\exists}
\RenewDocumentCommand{\exists}{}{\mathop{{}\ordinaryexists}}

\usepackage{tikz}
\usetikzlibrary{plotmarks,positioning,arrows}

\DeclareRobustCommand\full {\tikz[baseline=-0.6ex]\draw[very thick] (0,0)--(0.5,0);}
\DeclareRobustCommand\dotted{\tikz[baseline=-0.6ex]\draw[very thick,dotted] (0,0)--(0.5,0);}
\DeclareRobustCommand\dashed{\tikz[baseline=-0.6ex]\draw[very thick,dashed] (0,0)--(0.5,0);}
\DeclareRobustCommand\chain {\tikz[baseline=-0.6ex]\draw[very thick,dash dot] (0,0)--(0.5,0);}
\tikzset{%
    block/.style={draw, fill=white, rectangle, 
            minimum height=2em, minimum width=3em},
    input/.style={inner sep=0pt},       
    output/.style={inner sep=0pt},      
    sum/.style = {draw, fill=white, circle, minimum size=2mm, node distance=1.5cm, inner sep=0pt},
    pinstyle/.style = {pin edge={to-,thin,black}}
}

\definecolor{grey1}{rgb}{0.6,0.6,0.6}
\definecolor{yellow1}{rgb}{0.9290,0.6940,0.1250}
\definecolor{blue1}{rgb}{0,0.4470,0.7410}
\definecolor{red1}{rgb}{0.8500,0.3250,0.0980}
\definecolor{mag1}{rgb}{0.4940,0.1840,0.5560}
\definecolor{green1}{rgb}{0.4660,0.6740,0.1880}
\begin{document}
\title{An approach to the {LQG/LTR} design problem with specifications for finite-dimensional {SISO} control systems}
\author{Mahyar Mahinzaeim$^{\tt1,\ast}$}
\author{Kamyar Mehran$^{\tt2}$}

\thanks{
\vspace{-1em}\newline\noindent
{\sc Keywords}: {robust control systems design, LQG compensator, LTR technique, finite-dimensional SISO systems}
\newline\noindent
$^{\tt1}$ Research Center for Complex Systems, Aalen University, Germany.
\newline\noindent
 $^{\tt2}$ Antennas \& Electromagnetics Research Group, Queen Mary University of London, United Kingdom.
\newline\noindent
{\sc Emails}:
{\tt m.mahinzaeim@web.de},~{\tt k.mehran@qmul.ac.uk}.
\newline\noindent
$^{\ast}$ Corresponding author.
}

\begin{abstract}
This is an expository paper which discusses an approach to the linear quadratic Gaussian/loop transfer recovery (LQG/LTR) design problem for finite-dimensional single-variable (single-input/single-output, SISO) control systems. The approach is based on the utilisation of weighting augmentation for incorporating design specifications into the framework of the LTR technique for LQG compensator design. The LQG compensator is to simultaneously meet given analytical low- and high-frequency design specifications expressed in terms of desirable sensitivity and controller noise sensitivity functions. The paper is aimed at non-specialists and, in particular, practitioners in finite-dimensional LQG theory interested in the design of feedback compensators for closed-loop performance and robustness shaping of SISO control systems in realistic situations. The proposed approach is illustrated by a detailed design example: the torque control of a geared DC motor with an elastically mounted output shaft.
\end{abstract}
\maketitle

\pagestyle{myheadings} \thispagestyle{plain} \markboth{\sc M.\ Mahinzaeim, K.\ Mehran}{\sc {LQG/LTR} design with specifications for SISO systems}

\section{Introduction} 

The guaranteed performance (minimal sensitivity) properties of the Kalman--Bucy filter (KBF), and likewise the guaranteed robustness (gain and phase) margins of the linear quadratic regulator (LQR) state feedback controller, do not extend to the state-estimate feedback case, i.e.\ the linear quadratic Gaussian (LQG) compensator. This fact, which is well known in the control literature and goes back to a counterexample in \cite{Doyle1978}, led to a considerable amount of research in the 1980s to develop design methods that allow the robustness margins in state-estimate feedback control systems (in short \textit{feedback compensators}) to be ``restored''. This is generally considered to be the origin of the extensive development of robust control theory in the following decades, extending naturally classical single-variable (single-input/single-output, SISO) frequency or loop shaping problems to multi-variable (multi-input/multi-output, MIMO) design problems.

A well-established method for the systematic design of robust control systems is based on the so-called $H^\infty$ technique -- in fact any Hardy space norm (in $H^\infty$, $H^2$, etc.) control optimisation technique -- to allow for a frequency domain interpretation of \textit{design specifications}. The papers by Doyle and Stein \cite{DoyleStein1981} and Zames \cite{Zames1981} (in the same journal volume) are considered to be the forerunners of the theory, but it is only since the important work of Doyle et al.\ \cite{DoyleEtAl1989} that the $H^\infty$ technique can be specified in a suitable form for practical purposes in state space (earlier results in this direction can be found in \cite{Helton1976}). It is a fact, however, that the $H^\infty$ technique, despite, but more likely because of, its mathematically elegant form, and despite its maturity due to the numerous works that have been devoted to the method, has been only moderately successful in control engineering practise. In the books \cite{maciejowski1989multivariable,skogestad1996multivariable,Zhou1996,MR2080385,Burl1999,MorariZafiriou1989}, or for a more mathematical treatment see \cite{MR1353236,MR932459}, very informative insights into the method and its application in \textit{finite-dimensional} linear systems are given, along with several by now standard results in robust control theory (see also the important tutorial paper of Kwakernaak \cite{MR1211286}). Note that by ``finite-dimensional'' systems we mean systems governed by ordinary differential equations, leading to a finite-dimensional state space. In this paper we do not deal with infinite-dimensional systems, but in case the reader is interested, extensions of the $H^\infty$ design problem to a certain class of infinite-dimensional systems, including plants with distributed parameters and plants with time delays, are given in the monographs \cite{MR1269323,CurtainZwart1995,MR3914461} (a quick look through the control literature will reveal that research in infinite-dimensional contexts is still very much active in this area).

There are also formal approaches for LQG compensator design with (possibly arbitrarily) good performance or robustness, which are dual to each other. In fact, the hope that frequency-dependent design specifications for performance and robustness could also be accounted for in the design of LQG compensators was fulfilled earlier, for performance in the paper by Kwakernaak \cite{Kwakernaak1969}, and later also for robustness in the paper by Doyle and Stein \cite{DoyleStein1979}, with subsequent important additions and interpretations in \cite{24816,CHEN1991257,KAZEROONI01031986,DoyleStein1981,4788868,NtemannEtAl1991,LIU1990487,LehtomakiEtAl1981,4789039,4048281,SteinAthans1987,SafonovEtAl1981,1104723,MooreBlight1981,ZhangFreudenberg1990,Maciejowski1985,ShakedSoroka1985,BirdwellLaub1987,Prakash1990,doi:10.2514/3.20274,62280} to mention only a few. The result was the pragmatic development of the loop transfer recovery (LTR) technique for LQG compensator design, which quickly reached a high level of maturity even for MIMO control systems (the systems that require comparatively more involved processes for controller design). A general source on the LTR method is the monograph of Saberi et al.\ \cite{saberi1993loop}. (Much of the subject matter in the present paper is also covered in the less well-known 1981 AGARD/NATO lecture notes on MIMO control systems design and analysis \cite{MR4115834}, to which the reader may need to refer.)

As early as 1987, Dorato wrote in \cite{Dorato1987}: ``At the present time, one of the most commonly used techniques for robust design of multivariable systems appears to be the LQG/LTR technique.'' Today, in the robust control community, the LTR technique is commonly considered a mature one -- it is discussed, along with many examples, across all or at least most of the textbooks on optimal and robust control -- and consequently has somewhat faded as a research topic (although recent years have seen a number methodological advancements related to the use of LTR methods in general controller design, e.g., \cite{KumarEtAl2020,Lavretsky2012,https://doi.org/10.1049/iet-cta.2013.0671,CaliseYucelen2012,RavanbodEtAl2012,7064457,https://doi.org/10.1002/oca.2878,NiemannStoustrup2003,PaulaFerreira2011,daSilvaEtAl2014,8888190,ChenChen2008,IshiharaZheng2019,PereiraKienitz2014,IshiharaEtAl2005,MR4853336}). From the authors' point of view there are two reasons for this. The first reason is that in the last 30 to 40, there has been a general interest in doing research into infinite-dimensional control systems. In the process, completely new theories and methods have emerged, which have contributed a great deal to putting the control theory for infinite-dimensional linear systems on a solid mathematical basis (an earlier paper by the first author \cite{Mahinzaeim2019} is just one simple example of such a work), and there are now many books devoted to the subject, e.g., \cite{BensoussanEtAl2007,CurtainPritchard1978,CurtainZwart1995,MR1669395,MR2502023,MR4181480,Deutscher2012,Meurer2013}. To this end, it is not surprising that attempts have also been made in the late 1980s to generalise the LQG/LTR technique to infinite-dimensional control systems, see \cite{MatsonMaybeck1987,PaschallMaybeck1989,PaschallLair1989,MR945582}. 

The second important reason is that, around the same period, by far the largest part of the literature on robust control focussed on the Hardy space norm control optimisation methods as a means of specifying and ensuring the performance and robustness in LQG design. The first contributions to this were made, independently, by Bernstein and Haddad \cite{16419}, Grimble \cite{Grimble1989}, and Moore and Tay \cite{MooreTay1989}, and since the publication of these works numerous papers have been published dealing with various combinations of $H^\infty$, $H^2$ and LQG/LTR design methods, e.g., \cite{85062,MR1236784,StoustrupNiemann1993,Saeki1992,TuranMingori1995,daSilvaEtAl2014,PaulaFerreira2011,8888190,Grimble1990,ChristenEtAl1994,FerreiraGuaracy2021} and the references therein (for good textbook treatments, see, e.g., \cite{ColaneriEtAl1997,MR1669888}). It is interesting to note that the methods developed and illustrated in these papers could be of particular interest to the engineer wishing to obtain a trade-off between the norms in the design of feedback compensators based on Hardy space norm control optimisation techniques. It is also interesting to notice from the papers that the LQG/LTR design problem is a special case of the $H^\infty$ design problem and hence the Hardy space norm control optimisation techniques could, in principle, be used to solve any LQG design problem involving a finite-dimensional linear system. In general, however, this would first necessitate posing the control optimisation problem. This represents a disadvantage, along with the relatively high computational complexity in solving the control optimisation problem, which presents a drawback in the practical application of the procedure.

The majority of the work done on LQG/LTR design problems can be divided into two groups: either they concentrate purely on methodological advances and theoretical improvements, or they are restricted purely to application purposes. Both have their merits and need not be mutually exclusive; for a nice classical example, see the paper by Safonov et al.\ \cite{SafonovEtAl1981}. When looking at the vast application-related control literature, however, it quickly becomes apparent that the approaches used in many cases do not easily allow formal verification by the reader. In fact, their presentation is (certainly not always but) more often characterised by a series of imprecise arguments, which are made even more incomprehensible by the essentially heuristic nature of the LTR method due to the rather unsystematic adjustment procedure for the design parameters in the asymptotic recovery design step. For example, it is usually not at all obvious how the design parameters, of which there are four in the classical LQG theory, ``map'' into the resulting open-loop (transfer function) gain and phase characteristics. In addition, most of the work relates to very specific and complex MIMO control systems; classical examples are the tutorial papers \cite{PfeilEtAl1986,MartinEtAl1986,RodriguezAthans1986,BaumgartnerEtAl1986} presented at the 1986 American Control Conference and many others exist in the more recent control literature. As a consequence, the wealth of experience the control designer or practitioner can normally easily draw on for the purpose of applying a design method to finite-dimensional SISO control systems is comparatively small. It is important to mention in this connection that SISO control systems arise widely in practise (which accords with both the authors' experience in industry). Rarely, however, does the practitioner come across works in which an attempt is made to demonstrate the application of the LTR technique to LQG design problems for such systems and to provide a theoretically well-founded, yet easy to follow procedure for coping successfully with performance and robustness objectives. This paper is an attempt to provide such a work as an exposition on robust SISO control systems design based on the LQG/LTR technique. (The authors have been made aware of a relatively recent text by Hrycej \cite{Hrycej2018} in which a comprehensive application-driven treatment of robust SISO control systems design and analysis is given.)

Our aim here is to show how by reformulating the LQG design problem to incorporate frequency-dependent design specifications into the framework of the LTR technique, augmenting the nominal plant with appropriate weighting functions (in short \textit{weightings}), one will arrive at a completely algebraic procedure to find the LQG compensator transfer function; or equivalently the coefficients of the weightings will be determined solely via routine algebraic calculations to capture specific design goals or requirements for the LQG closed loop. Of course, formal weighting augmentation is not new in the robust control community -- in fact, one might say that it is the bread and butter of general robust control systems design -- and has been used or suggested to be used for the LQG/LTR design problem for a long time already, see \cite{DoyleStein1979,Doyle1978,SteinAthans1987,1104723,Athans1986x,PfeilEtAl1986,MartinEtAl1986,RodriguezAthans1986,MooreBlight1981}. Our approach to applying weighting augmentation, while not new, is different enough from those used in these papers as well as others that more generally apply augmentation procedures (see, e.g., \cite{Gupta1980,LoozeFreudenberg1988,MooreMingori1987,MooreXia1987}). An aspect -- a main shortcoming, in fact -- of the conventional LTR technique might explain this. It is well known that, in general, the full potential of the LQG/LTR technique will only be realised when full recovery of the guaranteed performance properties (respectively, robustness margins) of the KBF (respectively, LQR state feedback controller) is obtained. This happens in the case that the plant is minimum phase, i.e., if (and only if) the plant model has no zeros in the closed right half of the complex plane, which is essentially a consequence of Bode's classical integral relation (see the paper by Zhang and Freudenberg \cite{ZhangFreudenberg1990} for explanation, and in the more general context of feedback compensator designs, see, e.g., \cite{10.1093/imamci/2.2.153,467680}). However, if this feature is achieved, the open-loop gain and phase characteristics associated with the KBF or LQR state feedback controller will be retained in the final LQG compensator design and cause substantial problems in its practical implementation. This means, more concretely, the Nyquist plot of the LQG open-loop transfer function will remain outside a circle of unit radius centred at the familiar critical point in the complex plane, giving an infinite gain margin for the closed-loop system. Needless to say, no real system can maintain such a gain characteristic, noise being the troublesome factor, as discussed at length in \cite{HorowitzShaked1975}. Therefore, as set forth in many of the earlier papers quoted (also see \cite{Grimble1992,Fu1990,GerayLooze1996}) and as followed in other application-related works, one typically seeks only an approximate recovery in order to reduce the effect of noise and thereby to eliminate unnecessary high-frequency (i.e.\ beyond the crossover frequency) activity of the feedback controller. However, this in turn comes at the expense of a poorer low-frequency (i.e.\ below the crossover frequency) performance. It is precisely this simultaneous feedback compensator design to both low- and high-frequency design specifications which is generally the most challenging trade-off problem in any practical feedback controller design, and which has contributed to the motivation of the present work. The approach here, which, as far as the motivation is concerned, naturally has connections to other attempts, permits this to some degree and, in effect, presents an easy to follow procedure for systematic LQG compensator design with frequency-dependent design specifications.

We have structured the paper as follows. In Section \ref{sec02}, the state space descriptions of the nominal plant and weightings are introduced together with the corresponding transfer functions. By augmenting the weightings to the inputs of the nominal plant, we arrive at an augmented plant formulation in an extended or augmented state space so that it will be possible to incorporate frequency-dependent design specifications into the framework of the LTR technique for LQG compensator design. The motivation behind the use of (higher order) lead and lag type transfer functions as weightings is also discussed in Section \ref{sec02}. In Section \ref{sec03}, the LQG/LTR design problem with specifications is formulated wherein the design specifications are expressed analytically as upper bounds on the sensitivity and controller noise sensitivity magnitudes of the LQG closed loop in different frequency ranges. There, and in Section \ref{sec04}, we also summarise some of the key known results from classical LQG control theory and discuss their significance with respect to the LQG/LTR design problem. A straightforward and systematic technique is proposed for solving the problem, which leads to computable weightings. These weightings have the properties of defining loop shapings (made more precise in Section \ref{sec04}) and in Section \ref{sec05} are used to demonstrate the procedure by a detailed design example.

Finally, we note in passing that the goal of this work does not lie in the subtleties of the mathematical rigour of the results -- new or old -- simply because there are hardly any that have not already been discussed or proven elsewhere in the control literature. On the contrary, the purpose is to present some of the previous results in survey form, and, in particular, to emphasise their practical applicability. However, the reader will see that there are some interesting features and connections which have not been explored before and which, in fact, complement the known theoretical results in favour of practical applicability.

\section{Augmented plant formulation}\label{sec02}

Our plant is generic in the sense that it corresponds to the situation most typical of a practical (i.e., finite-dimensional) control system, namely when disturbances are collocated with the control inputs, the plant outputs are noisy, and the associated model transfer functions are rational and strictly proper or causal (i.e.\ the degree of the numerator of each transfer function is strictly less than that of the denominator). Consider the nominal plant model
\begin{equation}\label{eq13a}
\dot{x}_0=\bm{A}_0x_0+\bm{B}_0\left(r_1+r_2\right),\quad y=\bm{C}_0x_0+v
\end{equation}
with the state $x_0$ in the state space ${R}^n$. The variables $r_1$ and $r_2$ are designated as inputs, $y$ as output, and $v$ is noise. If the SISO representation is used, then there is only one nominal transfer function relating $r_1$ or $r_2$ to $y$ given by
\begin{equation*}\label{eq30}
{\mathbb{G}}_0\left(s\right)=\bm{C}_0 (s\bm{I}-\bm{A}_0)^{-1}\bm{B}_0.
\end{equation*}
The idea now is to augment the systems
\begin{alignat}{3}
 \dot{x}_j&=\bm{A}_jx_j+\bm{B}_j w_j,\quad&& r_j&&=\bm{C}_j x_j+\bm{D}_j w_j\label{eq14a},\quad j=1,2,
\end{alignat}
with $x_1$, $x_2$ in the spaces ${R}^p$ and ${R}^m$, respectively, to the inputs of the nominal plant model \eqref{eq13a} to obtain the augmented plant model
\begin{equation}\label{eq05a}
\dot{x}=\bm{A} x+\bm{B} w_1+\bm{F} w_2,\quad y=\bm{C} x+v
\end{equation}
in terms of the augmented state $x=(x_0,x_1,x_2)^\top$ in the augmented state space ${R}^n\times {R}^p\times {R}^m$, where
\begin{gather*}
\bm{A}=\left(\begin{matrix}
\bm{A}_0 & \bm{B}_0 \bm{C}_1 & \bm{B}_0 \bm{C}_2\\
\bm{0} & \bm{A}_1 &  \bm{0}\\
  \bm{0} &  \bm{0} & \bm{A}_2\\
\end{matrix}\right),\quad \bm{B}=\left(\begin{matrix}
\bm{B}_0 \bm{D}_1\\
\bm{B}_1\\
 \bm{0}
\end{matrix}\right),\quad \bm{F}=\left(\begin{matrix}
\bm{B}_0 \bm{D}_2 \\
 \bm{0} \\
\bm{B}_2
\end{matrix}\right),\quad
 \bm{C}=\left(\begin{matrix}
\bm{C}_0 &  \bm{0} &  \bm{0}
\end{matrix}\right);
\end{gather*}
we may assume without loss of generality the initial state is zero.

The matrices $\bm{A}_j$ in \eqref{eq14a} are assumed Hurwitz or stability matrices (i.e.\ they have only eigenvalues in the open left half of the complex plane). As can be seen directly, the set of eigenvalues of the matrix $\bm{A}$ in the form above is the unification of the sets of eigenvalues of the matrices $\bm{A}_0$, $\bm{A}_1$, and $\bm{A}_2$. Therefore, should $\bm{A}_0$ also be assumed to be a stability matrix, then $\bm{A}$ would be a stability matrix. This would have the important consequence that the well-known separation principle could be used for the LQG design (we shall return to this briefly in the next section).

The transfer functions of the systems \eqref{eq14a} are
\begin{equation*}
\mathcal{W}_j\left(s\right)=\bm{D}_j+\bm{C}_j (s{I}-\bm{A}_j)^{-1}\bm{B}_j,\quad j=1,2.
\end{equation*}
It is a simple consequence of standard matrix inversion techniques (see \cite{MR605440} for a concise overview of such techniques) that the transfer functions relating $w_1$ to $y$ and $w_2$ to $y$ in \eqref{eq05a} may then be given by
\begin{align*}
\bm{C} (s\bm{I}-\bm{A})^{-1}\bm{B}&=[\bm{D}_1+\bm{C}_1 (s\bm{I}-\bm{A}_1)^{-1}\bm{B}_1]\bm{C}_0 (s\bm{I}-\bm{A}_0)^{-1}\bm{B}_0\\
&=\mathcal{W}_1\left(s\right){\mathbb{G}}_0\left(s\right)
\end{align*}
and
\begin{align*}
\bm{C} (s\bm{I}-\bm{A})^{-1}\bm{F}&=[\bm{D}_2+\bm{C}_2 (s\bm{I}-\bm{A}_2)^{-1}\bm{B}_2]\bm{C}_0 (s\bm{I}-\bm{A}_0)^{-1}\bm{B}_0\\
&=\mathcal{W}_2\left(s\right){\mathbb{G}}_0\left(s\right),
\end{align*}
respectively. With appropriate care (the transfer function ${\mathbb{G}}_0\left(s\right)$ has so far not been assumed to be either stable or minimum phase) we can interpret $\mathcal{W}_1\left(s\right)$ and $\mathcal{W}_2\left(s\right)$ as the frequency-dependent weightings over any given frequency ranges for the designs of the LQR state feedback controller and the KBF, respectively. The transfer function block diagram of the augmented plant model is shown in Fig.\ \ref{fig01x}.

\begin{figure}
\centering
\resizebox{0.6\linewidth}{!}{\begin{tikzpicture}[auto, node distance=2cm, on grid, >=latex']
\node[input] (input1) {};
\node[input, above = of input1] (input2) {};
\node [block, right = of input1] (weight1) {$\mathcal{W}_1\left(s\right)$};
\node [sum, right = of weight1] (sum1) {\tiny$+$};
\node [block, right = of sum1] (plant1) {${\mathbb{G}}_0\left(s\right)$};
\node [sum, right = of plant1] (sum2) {\tiny$+$};
\node [output, right = of sum2] (output) {};
\node [block, above = of weight1] (weight2) {$\mathcal{W}_2\left(s\right)$};
\draw [draw,->] (input2) node[above right] (w2) {$w_2$} -- (weight2);
\draw [->] (plant1) -- node {} (sum2);
\draw [->] (sum1) -- node {} (plant1);
\draw [draw,<-] (sum2) -- ++(90:.6cm) node[above]{$v$};
\draw [draw,->] (input1) node[above right] {$w_1$} -- (weight1);
\draw [->] (weight1) -- node {} (sum1);
\draw [->] (weight2) -| node[] {} 
    node [near end] {} (sum1);
\draw [->] (sum2) -- (output) node [name=y, above left] {$y$};
\end{tikzpicture}\par}
\caption{Block diagram of augmented plant model}
\label{fig01x}
\end{figure}

We assume lead and lag type transfer functions of higher order for the weightings, given by
\begin{equation}\label{eqww324}
\mathcal{W}_1\left(s\right)=\frac{\tau^m_{12}}{\tau^m_{11}}\frac{(s+\tau_{11})^m}{(s+\tau_{12})^m},\quad \mathcal{W}_2\left(s\right)=\tau_{22}^p\frac{(s+\tau_{21})^p}{(s+\tau_{22})^p}, \quad m,p\ge 1,
\end{equation}
respectively, with the yet to be determined coefficients $\tau_{j1}$, $\tau_{j2}$, $j=1,2$, satisfying 
\begin{equation}\label{eqass01s}
0<\tau_{11}\le\tau_{12}<\infty,\quad 0<\tau_{22}\le\tau_{21}<\infty.
\end{equation}
The reader might object at this point that to obtain useful results (in the contexts here relevant) one should make stronger assumptions on the coefficients $\tau_{j1}$, $\tau_{j2}$ so that the pairs $\tau_{11}$, $\tau_{12}$ and $\tau_{21}$, $\tau_{22}$ are well separated from each other. Our design example in Section \ref{sec05} and the results preceding it demonstrate that this is indeed the case. (Note that for $\tau_{j1}=\tau_{j2}=1$, $j=1,2$, the weightings are just unity and the problem becomes that of nominal LQG design.)

\begin{figure}
\centering
\includegraphics[width=0.6\linewidth]{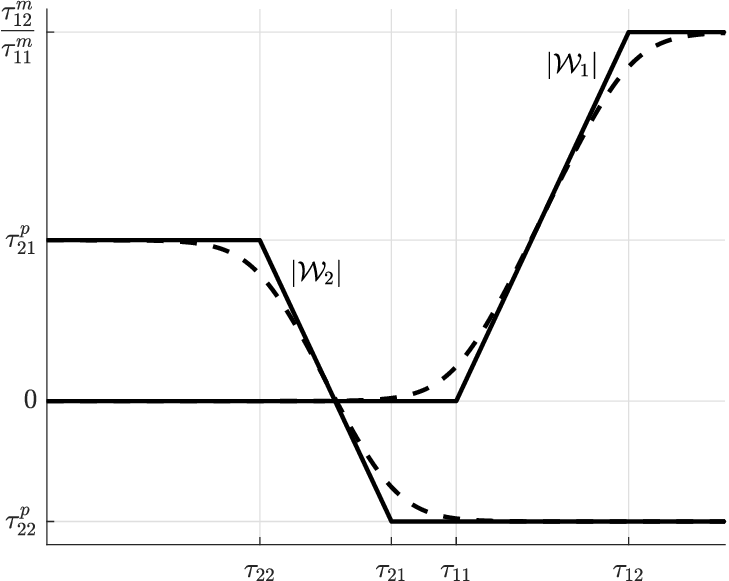}
\caption{Exact ($\dashed$) and asymptotic ($\full$) Bode frequency responses for weightings $\mathcal{W}_1\left(s\right)$ and $\mathcal{W}_2\left(s\right)$ given by \eqref{eqww324}}
\label{fig01}
\end{figure}

\begin{figure}
\centering
\includegraphics[width=0.6\linewidth]{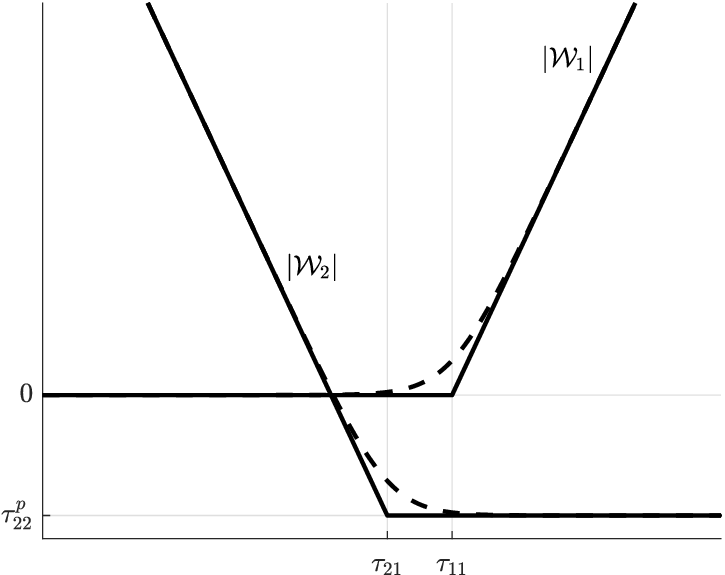}
\caption{Exact ($\dashed$) and asymptotic ($\full$) Bode frequency responses for weightings $\mathcal{W}_1\left(s\right)$ and $\mathcal{W}_2\left(s\right)$ given by \eqref{eqwwrr546}}
\label{fig01ss}
\end{figure}
There is, of course, a certain freedom in the choice of the weightings $\mathcal{W}_j\left(s\right)$. For example, other choices might be
\begin{equation}\label{eqwwrr546}
\mathcal{W}_1\left(s\right)=\frac{1}{\tau^m_{11}}({s+\tau_{11}})^m,\quad \mathcal{W}_2\left(s\right)=\tau_{22}^p\frac{(s+\tau_{21})^p}{s^p},\quad m,p\ge 1.
\end{equation}
In this case, as frequencies become small enough, $\mathcal{W}_2\left(s\right)$ becomes an ideal integrator of order $p$. On the other hand, as frequencies become large enough (noise frequencies), $\mathcal{W}_1\left(s\right)$ becomes an ideal differentiator of order $m$ and will eventually ensure the LQG open-loop gain has an asymptotic magnitude attenuation or roll-off of $\left(m+1\right)\times 20\,\text{dB}$ per decade. This is rather desirable from the practical point of view. However, the price to be paid for these choices is that $\mathcal{W}_1\left(s\right)$ and $\mathcal{W}_2\left(s\right)$ are now, respectively, improper and unstable, and hence some technical difficulties arise in the solution of the resulting LQG design problem. In particular, one needs to be careful not to suppose that the order $m$ of $\mathcal{W}_1\left(s\right)$ can generally be set arbitrarily high for the purpose of making the LQG open-loop gain to roll off faster. There is always a risk of loss of properness in $\mathcal{W}_1\left(s\right){\mathbb{G}}_0\left(s\right)$ when $m$ becomes too large; but again it is actually advantageous for noise attenuation to have $\mathcal{W}_1\left(s\right){\mathbb{G}}_0\left(s\right)$ improper and hence a strictly proper LQG compensator transfer function (see the remarks associated with \eqref{eq0ss8} in the next section).

Let us agree to restrict attention to the case where the weightings $\mathcal{W}_j\left(s\right)$ are defined as in \eqref{eqww324}. On a practical level, the choice of order of weightings is in general determined by low- and high-frequency design specifications, but may also be determined by at least equally important design specifications in and slightly above the crossover frequency range. There is, of course, always a trade-off between design specifications, the order of the feedback compensator, and the order of the plant model used in the design process. This is especially the case when the feedback compensator contains a higher order augmented plant model, as is the case here. Hence, as a practical requirement for lower order feedback compensator design, the restriction $m=p=1$ could be made in \eqref{eqww324} from the outset. As will be seen, this would be without loss of generality in our approach. However, we would like to keep the assumption that $m,p\ge 1$, noting that typically in practise, for a faster roll-off throughout the crossover frequency range and beyond, $m$ is chosen strictly larger than $p$ (this is indeed borne out by our design example).

The $\mathcal{W}_1$- and $\mathcal{W}_2$-gains are given by
\begin{equation}\label{eq12}
\left|\mathcal{W}_1\left(i\omega\right)\right|=\frac{\tau_{12}^m}{\tau_{11}^m}\frac{(\omega^2+\tau_{11}^2)^{m/2}}{(\omega^2+\tau_{12}^2)^{m/2} },\quad \left|\mathcal{W}_2\left(i\omega\right)\right|=\tau_{22}^p\frac{(\omega^2+\tau_{21}^2)^{p/2}}{(\omega^2+\tau_{22}^2)^{p/2}},\quad m,p\ge 1.
\end{equation}
The corresponding frequency responses over arbitrary frequency ranges are depicted in Fig.\ \ref{fig01}. For comparison purposes only, frequency responses are also shown in Fig.\ \ref{fig01ss} for the weightings in \eqref{eqwwrr546}. As indicated in Fig.\ \ref{fig01}, as $\omega\rightarrow 0$ and as $\omega\rightarrow \infty$, we find asymptotically that
\begin{equation*}
\left|\mathcal{W}_1\left(i\omega\right)\right|\simeq 1,\quad\left|\mathcal{W}_2\left(i\omega\right)\right|\simeq{\tau_{21}^p}\quad\text{and}\quad\left|\mathcal{W}_1\left(i\omega\right)\right|\simeq {\tau_{12}^m}/{\tau_{11}^m},\quad\left|\mathcal{W}_2\left(i\omega\right)\right|\simeq {\tau_{22}^p},
\end{equation*}
respectively. These properties show that by suitable choice of the pairs $\tau_{j1}$, $\tau_{j2}$, loop shapings are greatly simplified so that the design process becomes more practical. We expand on this in the sequel and make precise what we mean by ``suitable choice''.

In the next section the LQG/LTR design problem is formulated and the design specifications for it are carefully defined, after the relevant prerequisites have been established. The principal assumption, which we make explicit now, is that ${\mathbb{G}}_0\left(s\right)$ is minimum phase. We will see in Sections \ref{sec03} and \ref{sec04} that this assumption is important for our approach to the LQG/LTR design problem, and is not a technical convenience. (Recall from the remarks in the introduction that if one does not require full recovery, then the minimum-phase assumption is, in principle, not required and one can ``hope'' for sufficient recovery in the low frequency range where good performance is required.)

\section{LQG/LTR design problem formulation}\label{sec03}

We begin this section with a brief explanation of an underlying assumption in the LQG theory from a somewhat mathematical viewpoint. The LQG theory as it is commonly developed in engineering texts is based on the assumption that $w_2$ and $v$ in \eqref{eq05a} are independent white noise processes which are smooth, in the differentiable sense, as functions of time so that \eqref{eq05a} is well posed in the sense that it admits a unique classical solution. The smoothness hypothesis is in general false. For finite-dimensional linear systems (in fact also for infinite-dimensional systems), however, it can, at least to some weaker degree, be put on a firm mathematical basis -- even without the usual interpretation via stochastic integrals for Wiener processes -- if, at the most superficial level, one thinks in terms of white noise as the limit of a sequence of piecewise constant, temporally uncorrelated Gaussian random processes which become square integrable functions of the time variable on any finite interval; see the white noise formulation in \cite[Chapter VI]{Russell1979} (see also \cite{MR1122476}, and in \cite[Chapter 6]{Balakrishnan1981} a more general formulation, where the white noise is infinite-dimensional, is obtained). Then, for square integrable $w_1$, \eqref{eq05a} has a unique continuous solution with $y$ square integrable.

So let $w_2$, $v$ be independent white noise processes in the above sense, with zero means and unit variances, and therefore \eqref{eq05a} be well posed. For the LQG/LTR design problem the separation principle is most relevant. The associated results are well known for finite-dimensional linear systems and may be found in any standard textbook of optimal control and estimation theory, e.g., \cite{Russell1979,MR1020053,AndersonMoore1989,SpeyerChung2008,MR446628,KwakernaakSivan1972,Davis1977} (for more technical treatments, we refer the reader to \cite{Wonham1979,FlemingRishel1975}). We summarise the separation principle in a form more suitable to the present discussion and at the same time collect some (known and new) facts concerning the LQG/LTR design problem which we shall require. The KBF error process has the open-loop transfer function $\mathbb{M}\left(s\right)$ given by
\begin{equation}\label{eqMs}
\mathbb{M}\left(s\right)={\bm{C} (s\bm{I}-\bm{A})^{-1}\bm{S} \bm{C} ^\top}
\end{equation}
with the solution $\bm{S}=\bm{S}^\top$ to the so-called KBF matrix algebraic Riccati equation
\begin{equation}\label{eq07}
\bm{A} \bm{S}+\bm{S}\bm{A} ^\top-{\bm{S} \bm{C} ^\top \bm{C} \bm{S}}+\bm{F}\bm{F} ^\top=\bm{0}.
\end{equation}
The LQG open-loop transfer function for the nominal plant is
\begin{equation*}\label{eq09}
{\mathbb{G}}_0\left(s\right)\mathbb{K}\left(s\right)
\end{equation*}
with $\mathbb{K}\left(s\right)$ denoting the LQG compensator transfer function given by
\begin{equation}\label{eq06}
\mathbb{K}\left(s\right)={\rho}{\bm{B} ^\top \bm{P}} (s\bm{I}-\bm{A}+{\rho}{\bm{B}\bm{B} ^\top \bm{P}}+{\bm{S} \bm{C} ^\top \bm{C}})^{-1}{\bm{S} \bm{C} ^\top},
\end{equation}
with some constant $\rho>0$, and the solution $\bm{P}=\bm{P}^\top$ to the so-called LQR matrix algebraic Riccati equation
\begin{equation}\label{eq08}
\bm{A} ^\top\bm{P}+\bm{P}\bm{A}-{\rho}{\bm{P}\bm{B}\bm{B} ^\top\bm{P}}+\bm{C} ^\top \bm{C}=\bm{0}.
\end{equation}
The constant $\rho$ is the remaining design parameter, from the original four in the classical LQG theory, which is adjusted in the asymptotic recovery design step as follows. First notice that for the augmented plant model we can write
\begin{equation}\label{eq12fr}
\bm{C} (s\bm{I}-\bm{A})^{-1}\bm{B}\mathbb{K}\left(s\right)=\mathcal{W}_1\left(s\right){\mathbb{G}}_0\left(s\right)\mathbb{K}\left(s\right)=\mathbb{L}\left(s\right),
\end{equation}
where $\mathbb{L}\left(s\right)$ denotes the LQG open-loop transfer function for the augmented plant. Application now of a well-known matrix inversion result to \eqref{eq06} (cf.\ \cite[(16)]{MR605440}, wherein we identify $\bm{A}$ with $s\bm{I}-\bm{A}+{\bm{S} \bm{C} ^\top \bm{C}}$, $\bm{B}$ with $\bm{I}$, $\bm{U}$ with $\rho\bm{B}$, and $\bm{V}$ with $\bm{B}^\top\bm{P}$) shows, after some elementary algebra, that
\begin{equation*}
\mathbb{K}\left(s\right)=[1+{\rho}{\bm{B} ^\top \bm{P}} (s\bm{I}-\bm{A}+{\bm{S} \bm{C} ^\top \bm{C}})^{-1}\bm{B}{]}^{-1}{\rho}{\bm{B} ^\top \bm{P}} (s\bm{I}-\bm{A}+{\bm{S} \bm{C} ^\top \bm{C}})^{-1}{\bm{S} \bm{C} ^\top}.
\end{equation*}
Equivalently, making a trivial factorisation of $\rho$,
\begin{equation*}
\mathbb{K}\left(s\right)=[\rho^{-1/2}+\rho^{1/2}{\bm{B} ^\top \bm{P}} (s\bm{I}-\bm{A}+{\bm{S} \bm{C} ^\top \bm{C}})^{-1}\bm{B}{]}^{-1}{\rho}^{1/2}{\bm{B} ^\top \bm{P}} (s\bm{I}-\bm{A}+{\bm{S} \bm{C} ^\top \bm{C}})^{-1}{\bm{S} \bm{C} ^\top}.
\end{equation*}
If we now consider \eqref{eq08} with $\rho$ large enough so that, since ${\mathbb{G}}_0\left(s\right)$ by assumption is minimum phase, and thus also is $\mathcal{W}_1\left(s\right){\mathbb{G}}_0\left(s\right)$, $\bm{P}$ is close to the zero matrix (first proven in the paper by Kwakernaak and Sivan \cite{KwakernaakSivan1972a}), asymptotically, as $\rho\rightarrow \infty$, we obtain
\begin{equation*}
{\rho}{\bm{P}\bm{B}\bm{B} ^\top\bm{P}}=({\rho}^{1/2}\bm{B}^\top\bm{P})^\top{\rho}^{1/2}\bm{B} ^\top\bm{P}\simeq\bm{C} ^\top \bm{C},
\end{equation*}
and therefore
\begin{equation}\label{eq0ss8dd}
\mathbb{K}\left(s\right)\simeq[\bm{C} (s\bm{I}-\bm{A}+{\bm{S} \bm{C} ^\top \bm{C}})^{-1}\bm{B}{]}^{-1}\bm{C} (s\bm{I}-\bm{A}+{\bm{S} \bm{C} ^\top \bm{C}})^{-1}{\bm{S} \bm{C} ^\top}.
\end{equation}
Now, using the matrix inversion result mentioned above again, with some elementary algebra, we have
\begin{align*}
\bm{C} (s\bm{I}-\bm{A}+{\bm{S} \bm{C} ^\top \bm{C}})^{-1}\bm{B}&=[1+\bm{C} (s\bm{I}-\bm{A})^{-1}{\bm{S} \bm{C} ^\top}{]}^{-1}\bm{C} (s\bm{I}-\bm{A})^{-1}\bm{B}\\
&=(1+\mathbb{M}\left(s\right))^{-1}\mathcal{W}_1\left(s\right){\mathbb{G}}_0\left(s\right);
\end{align*}
a similar calculation shows that
\begin{align*}
\bm{C} (s\bm{I}-\bm{A}+{\bm{S} \bm{C} ^\top \bm{C}})^{-1}{\bm{S} \bm{C} ^\top}&=[1+\bm{C} (s\bm{I}-\bm{A})^{-1}{\bm{S} \bm{C} ^\top}{]}^{-1}\bm{C} (s\bm{I}-\bm{A})^{-1}{\bm{S} \bm{C} ^\top}\\
&=(1+\mathbb{M}\left(s\right))^{-1}\mathbb{M}\left(s\right).
\end{align*}
Substituting these expressions in \eqref{eq0ss8dd} we find
\begin{equation}\label{eq0ss8}
\mathbb{K}\left(s\right)\simeq[(1+\mathbb{M}\left(s\right))^{-1}\mathcal{W}_1\left(s\right){\mathbb{G}}_0\left(s\right){]}^{-1}(1+\mathbb{M}\left(s\right))^{-1}\mathbb{M}\left(s\right)=(\mathcal{W}_1\left(s\right){\mathbb{G}}_0\left(s\right))^{-1}\mathbb{M}\left(s\right).
\end{equation}
Returning to \eqref{eq12fr}, and substituting for $\mathbb{K}\left(s\right)$ from \eqref{eq0ss8}, the conclusion is that \eqref{eq12fr} takes the asymptotic form, as $\rho\rightarrow \infty$,
\begin{equation}\label{eq12r4}
\mathbb{L}\left(s\right)\simeq \mathbb{M}\left(s\right)
\end{equation}
for all $s=i\omega$ with $\omega\in{R}$. It is then that asymptotic recovery of the guaranteed performance properties of the KBF is said to occur, and again the reason for the minimum-phase assumption is now also clear from \eqref{eq0ss8} (i.e., the inversion of $\mathcal{W}_1\left(s\right){\mathbb{G}}_0\left(s\right)$). More details of such calculations leading to \eqref{eq12r4} can be found in many places, e.g., \cite[Section 8.4]{AndersonMoore1989}, \cite[Section 5.4]{maciejowski1989multivariable}, or \cite[Section 9.9.2]{SpeyerChung2008} and in almost any of the papers cited in the introduction (they seem to have first appeared in the appendix of the paper by Stein and Athans \cite{SteinAthans1987}).

In essence, we have shown above that \eqref{eq12r4} provides a basis for a complete solution to the LQG/LTR design problem. To further substantiate this, we must show that in the frequency ranges in which the design specifications will be given effective manipulations may be made via the weightings $\mathcal{W}_j\left(s\right)$. For this purpose we recall that the sensitivity function and controller noise sensitivity function of the LQG closed loop are
\begin{equation*}
\mathbb{S}\left(s\right)=(1+\mathbb{L}\left(s\right))^{-1},\quad \mathbb{K}\left(s\right)\mathbb{S}\left(s\right),
\end{equation*}
respectively. Recall that if $\bm{A}$ is a stability matrix, which implies that there exist unique non-negative symmetric solutions $\bm{S}$, $\bm{P}$ to \eqref{eq07} and \eqref{eq08}, respectively, $\mathbb{S}\left(i\omega\right)$ and $\mathbb{K}\left(i\omega\right)\mathbb{S}\left(i\omega\right)$ are well defined for all $\omega\in {R}$. So the conditions in Section \ref{sec02} that $\bm{A}_1$ and $\bm{A}_2$ be stability matrices cannot be dropped in the above considerations if we assume stabilisability to hold for the open-loop systems (i.e., the minimum requirements for existence of non-negative symmetric solutions to \eqref{eq07} and \eqref{eq08}).

Finally, we note for use later in Section \ref{sec05} that the nominal sensitivity function is denoted by $\mathbb{S}_0\left(s\right)$, so that
\begin{equation*}
\mathbb{S}_0\left(s\right)=(1+{\mathbb{G}}_0\left(s\right)\mathbb{K}\left(s\right))^{-1}.
\end{equation*}

\subsection{Design specifications}

To simplify the notation, we will use throughout the paper $\omega_0$ to denote any of the unity gain crossover frequencies for the open-loop systems and define the frequency regions
\begin{equation*}
\mathit{\Delta}^{\omega_0}=\left\{\omega_{11},\omega_{12}\in \left(0,\omega_0\right):\omega_{11}<\omega_{12}\right\},\quad \mathit{\Delta}_{\omega_0}=\left\{\omega_{21},\omega_{22}\in \left(\omega_0,\infty\right):\omega_{21}<\omega_{22}\right\}.
\end{equation*}
Here $\omega_{1k}$, $\omega_{2k}$, $k=1,2$, denote the prescribed frequencies at which we wish to enforce design specifications. The LQG/LTR design problem considered is then to find $\mathbb{K}\left(s\right)$ in the form \eqref{eq06} such that, for prescribed bounds $m_{1k}$, $m_{2k}$, $k=1,2$, the resulting closed loop meets the following requirements in $\mathit{\Delta}^{\omega_0}$ and $\mathit{\Delta}_{\omega_0}$, respectively:
\begin{itemize}[leftmargin=*,align=left,labelwidth=\parindent]
\item {Tracking and disturbance rejection:}
\begin{equation}\label{eq_04}
\left|\mathbb{S}\left(i\omega_{1k}\right)\right|\leq m_{1k},\quad k=1,2.
\end{equation}
\item {Controller noise attenuation:}
\begin{equation}\label{eq_0s4}
\left|\mathbb{K}\left(i\omega_{2k}\right)\mathbb{S}\left(i\omega_{2k}\right)\right|\leq m_{2k},\quad k=1,2.
\end{equation}
\end{itemize}

The design specifications \eqref{eq_04} and \eqref{eq_0s4} are the low- and high-frequency design specifications, respectively, and, as discussed in the introduction, cannot normally be expected to be achieved automatically using the conventional LTR technique together with the nominal plant. It should be clear that with the requirements in \eqref{eq_0s4} and \eqref{eq_04}, we ultimately wish to prevent, respectively, excessive control activity in $\mathit{\Delta}_{\omega_0}$ and yet maintain a satisfactory degree of performance in $\mathit{\Delta}^{\omega_0}$. The bound pairs $m_{1k}$, $m_{2k}$ usually are fixed throughout the design process.

In the approach here, it is the shaping of the weightings $\mathcal{W}_j\left(s\right)$ which is important and which is exploited in the frequency regions referred to above to achieve the design specifications. However, the design specifications \eqref{eq_04} and \eqref{eq_0s4} are technically somewhat difficult to handle as they stand. The next objective is thus clear: translate the closed-loop requirements set out in \eqref{eq_04} and in \eqref{eq_0s4} into requirements, possibly for the open loop, involving the weightings. This is explored in the following section.

\section{Solution of the LQG/LTR design problem}\label{sec04}

The objective in this section is to formalise and demonstrate our approach to obtaining a solution to the LQG/LTR design problem considered. Simple arguments will show that to guarantee simultaneous adherence to the requirements in \eqref{eq_04} and \eqref{eq_0s4} it will be possible in the corresponding frequency regions $\mathit{\Delta}^{\omega_0}$ and $\mathit{\Delta}_{\omega_0}$ to shape the $\mathcal{W}_2$- and $\mathcal{W}_1$-gains. In order to derive the approach, we need some basic preliminary notions, the most important of which are the magnitude approximations of $\left|\mathbb{L}\left(i\omega\right)\right|$ in $\mathit{\Delta}^{\omega_0}$ and $\mathit{\Delta}_{\omega_0}$. (For more details on what follows, it is enough to consult the survey paper by Freudenberg et al.\ \cite{FreudenbergEtAl2000}.)

It is not difficult to see from the triangle inequality that for all $\omega\in {R}$ the estimate $\left|\mathbb{L}\left(i\omega\right)\right|-1\leq{\left|1+\mathbb{L}\left(i\omega\right)\right|}\leq\left|\mathbb{L}\left(i\omega\right)\right|+1$ is valid, from which, since $\mathbb{S}\left(i\omega\right)=(1+\mathbb{L}\left(i\omega\right))^{-1}$, an estimate of the form
\begin{equation*}
\left|\mathbb{L}\left(i\omega\right)\right|-1\leq{\left|\mathbb{S}\left(i\omega\right)\right|^{-1}}\leq\left|\mathbb{L}\left(i\omega\right)\right|+1
\end{equation*}
follows. If in $\mathit{\Delta}^{\omega_0}$ we have $\left|\mathbb{L}\left(i\omega\right)\right|$ large enough so that, by the above estimate, $\left|\mathbb{S}\left(i\omega\right)\right|$ is approximated by $\left|\mathbb{S}\left(i\omega\right)\right|\simeq {\left|\mathbb{L}\left(i\omega\right)\right|}^{-1}$, 
then we may use, instead of \eqref{eq_04},
\begin{equation}\label{eqL01a}
\left|\mathbb{L}\left(i\omega_{1k}\right)\right|\leq \frac{1}{m_{1k}},\quad k=1,2.
\end{equation}
On the other hand, if in $\mathit{\Delta}_{\omega_0}$ we have $\left|\mathbb{L}\left(i\omega\right)\right|$ small enough so that $\left|\mathbb{S}\left(i\omega\right)\right|\simeq 1$, then \eqref{eq_0s4} becomes
\begin{equation}\label{eqaL01ssa}
\left|\mathbb{K}\left(i\omega_{2k}\right)\right|\leq m_{2k},\quad k=1,2.
\end{equation}
It follows immediately from \eqref{eq12fr} and the trivial identity $\left|\mathbb{L}\left(i\omega\right)\right|=\left|\mathcal{W}_1\left(i\omega\right){\mathbb{G}}_0\left(i\omega\right)\mathbb{K}\left(i\omega\right)\right|=\left|\mathcal{W}_1\left(i\omega\right)\right|\left|{\mathbb{G}}_0\left(i\omega\right)\right|\left|\mathbb{K}\left(i\omega\right)\right|$ that we may then use
\begin{equation}\label{eqL01b}
\left|\mathcal{W}_1\left(i\omega_{2k}\right)\right|\ge \frac{1}{m_{2k}}\left|\mathbb{L}\left(i\omega_{2k}\right)\right|\left|{\mathbb{G}}_0\left(i\omega_{2k}\right)\right|^{-1},\quad k=1,2,
\end{equation}
instead of \eqref{eqaL01ssa}. In Section \ref{sec03} it has been justified to be correct by the conclusion \eqref{eq12r4} to replace $\mathbb{L}\left(i\omega\right)$ by $\mathbb{M}\left(i\omega\right)$ for $\rho\rightarrow \infty$ and for all $\omega\in {R}$. Let us consider this situation and write \eqref{eqL01a} as
\begin{equation}\label{eqL0xx1a}
\left|{\mathbb{M}\left(i\omega_{1k}\right)}\right|\leq \frac{1}{m_{1k}},\quad k=1,2.
\end{equation}
We can now simplify the situation even further. However, before going on to this, it is useful to recall, very briefly, what is meant when we say the performance properties of the KBF (or the robustness margins of the LQR state feedback controller in the dual problem) are ``guaranteed''. To this end, we rewrite the KBF matrix algebraic Riccati equation \eqref{eq07} as
\begin{equation*}
\bm{S} (-s\bm{I}-\bm{A}) ^\top+\left(s\bm{I}-\bm{A}\right) \bm{S}+{\bm{S} \bm{C} ^\top \bm{C} \bm{S}}=\bm{F}\bm{F} ^\top.
\end{equation*}
Pre- and post-multiply by $\bm{C} (s\bm{I}-\bm{A})^{-1} $ and $[\bm{C} (-s\bm{I}-\bm{A})^{-1}]^\top$, respectively, and add $1$ to both sides of the resulting expression. With rearranging, this gives
\begin{equation*}
[1+{\bm{C} (-s\bm{I}-\bm{A})^{-1}\bm{S} \bm{C} ^\top}][1+{\bm{C} (s\bm{I}-\bm{A})^{-1}\bm{S} \bm{C} ^\top}]=1+[\bm{C} (-s\bm{I}-\bm{A})^{-1}\bm{F}][\bm{C} (s\bm{I}-\bm{A})^{-1}\bm{F}].
\end{equation*}
Then putting in $s=i\omega$, for any $\omega\in {R}$, it is easily verified that
\begin{equation*}
|1+{\bm{C} (i\omega\bm{I}-\bm{A})^{-1}\bm{S} \bm{C} ^\top}|^2=1+|\bm{C} (i\omega\bm{I}-\bm{A})^{-1}\bm{F}|^2,
\end{equation*}
and thus
\begin{equation}\label{eqineqk1}
\left|1+\mathbb{M}\left(i\omega\right)\right|^2={1+{\left|\mathcal{W}_2\left(i\omega\right){\mathbb{G}}_0\left(i\omega\right)\right|^2}}.
\end{equation}
Clearly the interpretation to be given to \eqref{eqineqk1} is then $\left|1+\mathbb{M}\left(i\omega\right)\right|\ge 1$. This inequality shows that, for all $\omega\in {R}$, the sensitivity function gain associated with the KBF, which is just the reciprocal of $\left|1+\mathbb{M}\left(i\omega\right)\right|$, is bounded with bound unity; or, equivalently, the Nyquist plot of $\mathbb{M}\left(s\right)$ always avoids the unit circle centred at the critical point $-1+0i$ in the complex plane (the result was first discovered for the dual problem by Kalman \cite{Kalman1964x}).

Now, just as before, an application of the magnitude approximation of $\left|\mathbb{M}\left(i\omega\right)\right|$ in $\mathit{\Delta}^{\omega_0}$ -- i.e.\ $\left|\mathbb{M}\left(i\omega\right)\right|$ is large enough in $\mathit{\Delta}^{\omega_0}$ -- yields using \eqref{eqineqk1} that $\left|\mathbb{M}\left(i\omega\right)\right|\simeq{\left|\mathcal{W}_2\left(i\omega\right){\mathbb{G}}_0\left(i\omega\right)\right|}= \left|\mathcal{W}_2\left(i\omega\right)\right|\left|{\mathbb{G}}_0\left(i\omega\right)\right|$, which we can use in \eqref{eqL0xx1a}, finally obtaining
\begin{equation}\label{aassqw}
\left|{\mathcal{W}_2\left(i\omega_{1k}\right)}\right|\leq \frac{1}{m_{1k}}\left|{{\mathbb{G}}_0\left(i\omega_{1k}\right)}\right|^{-1},\quad k=1,2.
\end{equation}
Moreover, as a consequence of the limiting case $\rho\rightarrow \infty$, \eqref{eqL01b} becomes
\begin{equation}\label{aasssqw}
\left|\mathcal{W}_1\left(i\omega_{2k}\right)\right|\ge \frac{1}{m_{2k}}\left|\mathbb{M}\left(i\omega_{2k}\right)\right|\left|{\mathbb{G}}_0\left(i\omega_{2k}\right)\right|^{-1},\quad k=1,2.
\end{equation}
To summarise, instead of the closed-loop requirements in \eqref{eq_04} and \eqref{eq_0s4}, we can now use the corresponding open-loop requirements in \eqref{aassqw} and \eqref{aasssqw}. Using the expressions for the $\mathcal{W}_1$- and $\mathcal{W}_2$-gains given by \eqref{eq12} in \eqref{aasssqw} and \eqref{aassqw}, respectively, we are led to the conclusion that a system of non-linear algebraic inequalities on $\left|\mathcal{W}_1\left(i\omega_{2k}\right)\right|$, $\left|\mathcal{W}_2\left(i\omega_{1k}\right)\right|$, $k=1,2$, with respect to the unknowns $\tau_{j1}$, $\tau_{j2}$, $j=1,2$, results. The inequalities are coupled together through the dependence of $\mathcal{W}_2\left(s\right)$, and hence $\mathbb{M}\left(s\right)$, on the pair $\tau_{21}$, $\tau_{22}$. This is, of course, exactly what we want for simultaneous shaping of the weighting $\mathcal{W}_2\left(s\right)$ in $\mathit{\Delta}^{\omega_0}$ and $\mathcal{W}_1\left(s\right)$ in $\mathit{\Delta}_{\omega_0}$, and thus simultaneous LQG compensator design to both low- and high-frequency design specifications.

The problem of finding the LQG compensator transfer function $\mathbb{K}\left(s\right)$ thus reduces to that of finding suitable solution pairs $\tau_{j1}$, $\tau_{j2}$ for prescribed bound pairs $m_{1k}$, $m_{2k}$, and for this, we must make precise what suitable solution pairs are. It is clear from what has been said in Section \ref{sec02} that if solution pairs $\tau_{j1}$, $\tau_{j2}$ exist, then they should satisfy the conditions in \eqref{eqass01s}. It is also clear that this alone is not enough to guarantee simultaneous adherence to the requirements in \eqref{aassqw} and \eqref{aasssqw}, for it might well happen that the intervals $0<\tau_{22}\le\tau_{21}<\infty$ and $0<\tau_{11}\le\tau_{12}<\infty$ overlap without changing the $\mathbb{M}$-gain (i.e.\ we might no longer have the conclusion \eqref{eq12r4} in the region $\mathit{\Delta}^{\omega_0}$). So, to ensure that the high-frequency requirements in \eqref{aasssqw} are met without deterioration in the low-frequency performance requirements in \eqref{aassqw}, it is necessary to assume that the intervals are well separated, as anticipated in the paragraph following \eqref{eqass01s}. The requirements in \eqref{aassqw} and \eqref{aasssqw} are specified below and beyond the crossover frequency, respectively, so it is reasonable to ``strengthen'' the assumptions \eqref{eqass01s} to
\begin{equation}\label{eqass01sx}
0<\tau_{22}\le\tau_{21}<\omega_0,\quad \omega_0<\tau_{11}\le\tau_{12}<\infty,
\end{equation}
and look for solution pairs $\tau_{j1}$, $\tau_{j2}$ that satisfy these conditions. 

It is important to point out that the regions of existence of suitable solution pairs $\tau_{j1}$, $\tau_{j2}$ vary drastically with the bound pairs $m_{1k}$, $m_{2k}$. Indeed, solution pairs may not exist at all, and hence the design specifications may not be achievable, for certain values of the bounds. This highlights the general problem of attainability, i.e., the problem of the achievable performance and robustness of a feedback compensator design, and is indicative of the iterative nature of any control design process. If solution pairs do exist, then clearly their values are used to define the weightings $\mathcal{W}_j\left(s\right)$, and ultimately $\mathbb{K}\left(s\right)$ can be determined through these, as has all been explained.

The values of the pairs $\tau_{j1}$, $\tau_{j2}$ can be obtained from an iterative solution routine. However, returning to the problem of attainability, this may prove difficult numerically for the inequality case and computation on the boundaries of \eqref{aassqw}, \eqref{aasssqw} might be a better approach; it is very common in practise that design specifications are given as equality constraints. The following example gives insight into the details of the design process when this situation arises. All the calculations are standard ones and can be performed using Matlab or Maple.

\section{Design example}\label{sec05}

In this section, an example of LQG compensator design using the previously described design process is considered. The example deals with the design of a SISO control system for a geared DC motor with an elastically mounted output shaft. This system finds application in many practical electromechanical systems where conversion of electromagnetic energy into mechanical energy takes place. These particularity include systems which require high-precision performance of a feedback compensator, e.g., power steering systems, adjustable steering wheels, and robot manipulators.

\begin{figure}
\centering
\includegraphics[width=0.6\linewidth]{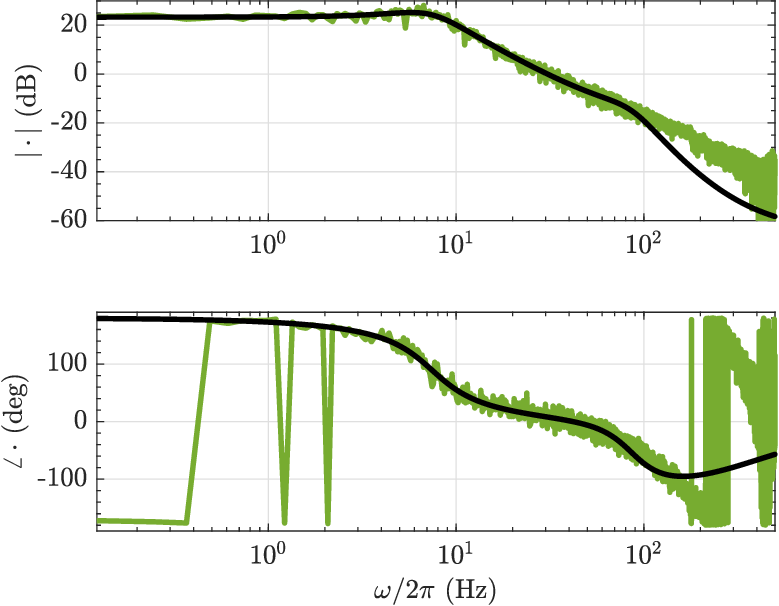}
\caption{Plant model (${\color{black}\full}$) and measured (${\color{green1}\full}$) frequency responses}
\label{fig02}
\end{figure}

The DC motor acts as a servo, driving the shaft torque to zero. The motor torque is considered to be the plant control input and the shaft torque is the plant output. The DC motor is current controlled and we assume a perfect controller in the sense that the control inputs are assumed to perfectly match the implemented motor torques. The nominal plant model is determined experimentally over a frequency range by measuring the shaft torque due to exciting motor torques of various amplitudes. The transfer function specifying the relationship between these quantities is calculated to be
\begin{equation}\label{dds3wee}
{\mathbb{G}}_0\left(s\right)=\frac{-2.06868s^3-8000.94442s^2-1.35637\times 10^7s-9.25775\times 10^9}{s^4+414.17923s^3+3.05819\times 10^5s^2+1.30484\times 10^7s+6.33087\times 10^8},
\end{equation}
and is stable and minimum phase (it is obviously rational and strictly proper). The corresponding frequency response is shown in Fig.\ \ref{fig02}, along with the mean measured frequency response of the plant. It is seen that the nominal plant model ${\mathbb{G}}_0\left(s\right)$ given by \eqref{dds3wee} matches very well with the measured frequency response up to a frequency of about $80\times2\pi\,\text{rad}/\text{sec}$. This is considered satisfactory for the purpose of achieving the low-frequency design specifications \eqref{eq_04} via shaping of the weighting $\mathcal{W}_2\left(s\right)$ to satisfy the open-loop requirements in \eqref{aassqw}; the mismatch at frequencies above $80\times2\pi\,\text{rad}/\text{sec}$ is addressed by shaping $\mathcal{W}_1\left(s\right)$ to satisfy the open-loop requirements in \eqref{aasssqw} and thereby to achieve the high-frequency design specifications \eqref{eq_0s4}. It is also seen that the phase of the plant frequency response has a characteristic that rules out simple ``mass-spring-damper'' oscillatory behaviour which requires that the phase tends to a constant value at high frequencies. Mechanically such a characteristic may be indicative of backlash in the gear.

\begin{figure}
\centering
\begin{subfigure}[t]{\linewidth}
\centering\includegraphics[width=0.6\linewidth]{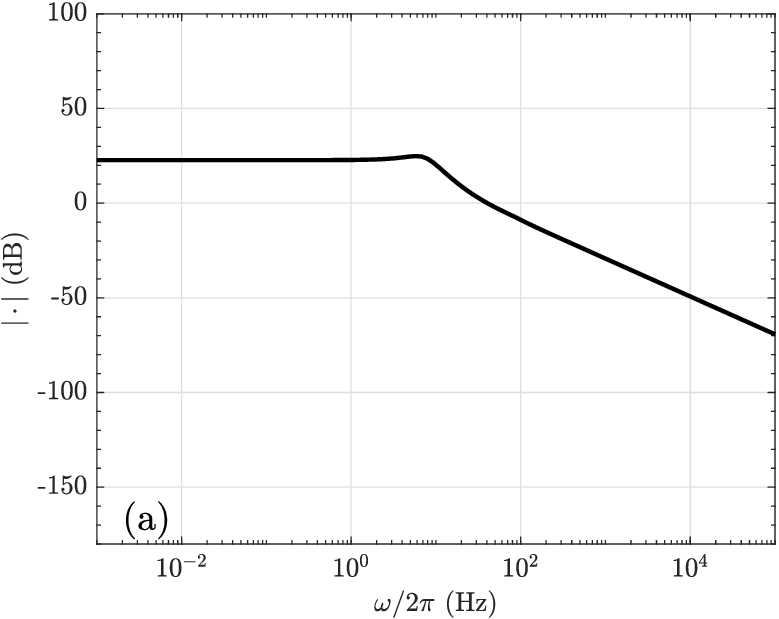}
\end{subfigure}
\medskip
\begin{subfigure}[t]{\linewidth}
\centering\includegraphics[width=0.6\linewidth]{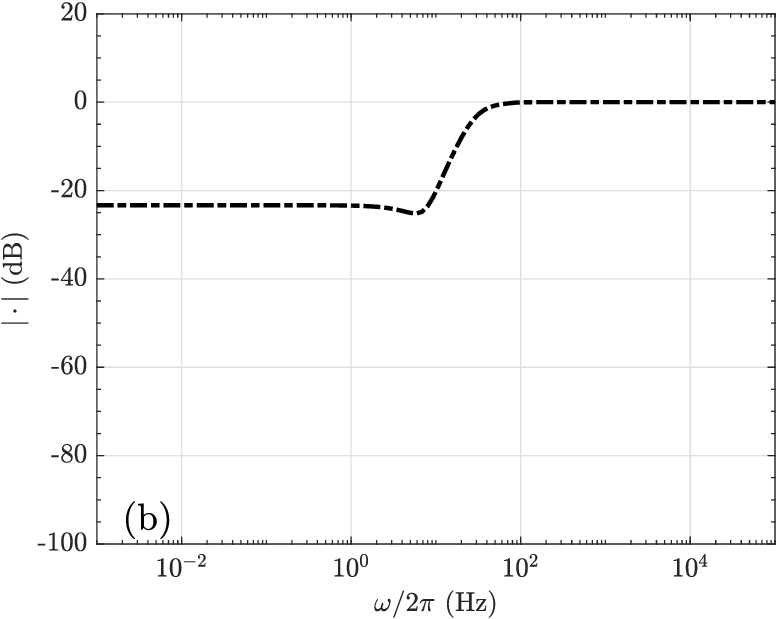}
\end{subfigure}
\caption{Bode frequency responses for (a) $\mathbb{M}_0\left(s\right)$ (${\color{black}\full}$); (b) $(1+\mathbb{M}_0\left(s\right))^{-1}$ (${\color{black}\chain}$)}\label{fig03}
\end{figure}

\begin{figure}
\centering
\begin{subfigure}[t]{\linewidth}
\centering\includegraphics[width=0.6\linewidth]{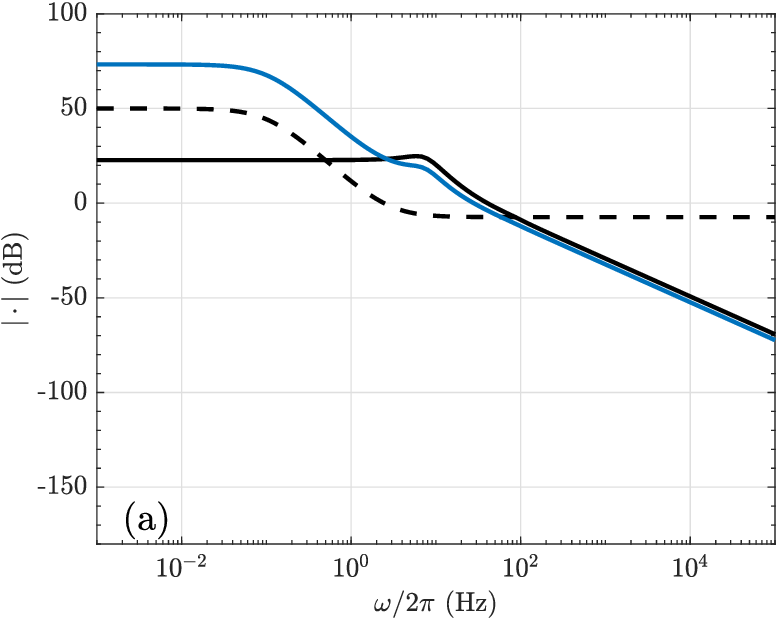}
\end{subfigure}
\medskip
\begin{subfigure}[t]{\linewidth}
\centering\includegraphics[width=0.6\linewidth]{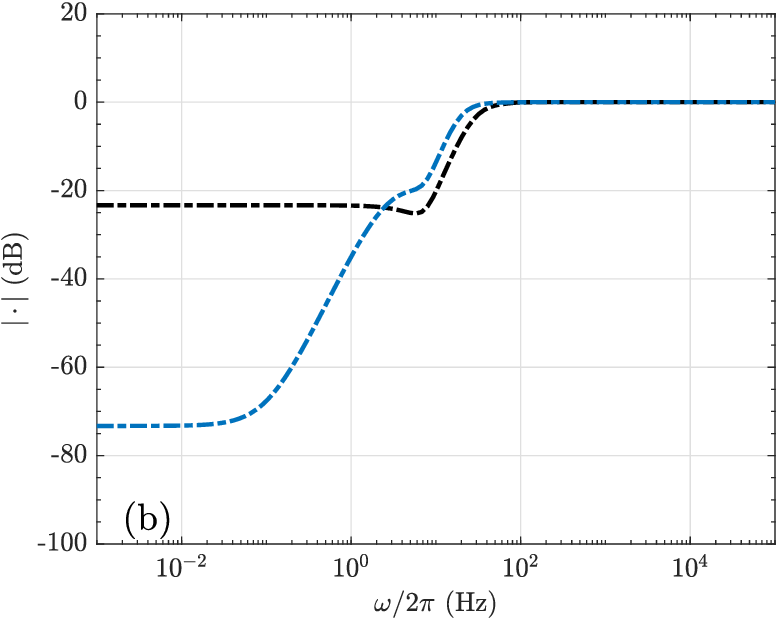}
\end{subfigure}
\caption{Bode frequency responses for (a) $\mathbb{M}_0\left(s\right)$ (${\color{black}\full}$), $\mathbb{M}\left(s\right)$ (${\color{blue1}\full}$), and $\mathcal{W}_2\left(s\right)$ (${\color{black}\dashed}$); (b) $(1+\mathbb{M}_0\left(s\right))^{-1}$ (${\color{black}\chain}$) and $(1+\mathbb{M}\left(s\right))^{-1}$ (${\color{blue1}\chain}$)}\label{fig04}
\end{figure}

\begin{figure}
\centering
\begin{subfigure}[t]{\linewidth}
\centering\includegraphics[width=0.6\linewidth]{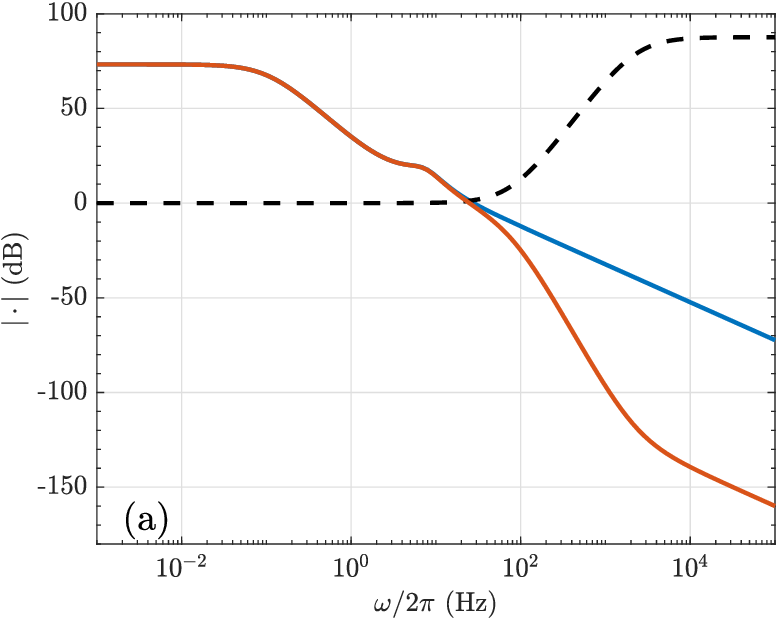}
\end{subfigure}
\medskip
\begin{subfigure}[t]{\linewidth}
\centering\includegraphics[width=0.6\linewidth]{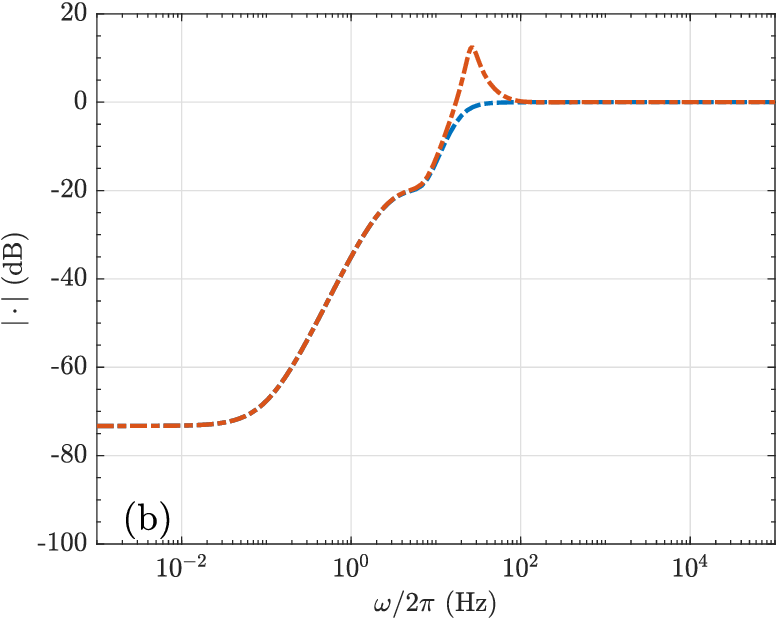}
\end{subfigure}
\caption{Bode frequency responses for (a) $\mathbb{M}\left(s\right)$ (${\color{blue1}\full}$), ${\mathbb{G}}_0\left(s\right)\mathbb{K}\left(s\right)$ (${\color{red1}\full}$), and $\mathcal{W}_1\left(s\right)$ (${\color{black}\dashed}$); (b) $(1+\mathbb{M}\left(s\right))^{-1}$ (${\color{blue1}\chain}$), $\mathbb{S}_0\left(s\right)$ (${\color{red1}\chain}$)}\label{fig05}
\end{figure}

\begin{figure}
\centering
\includegraphics[width=0.6\linewidth]{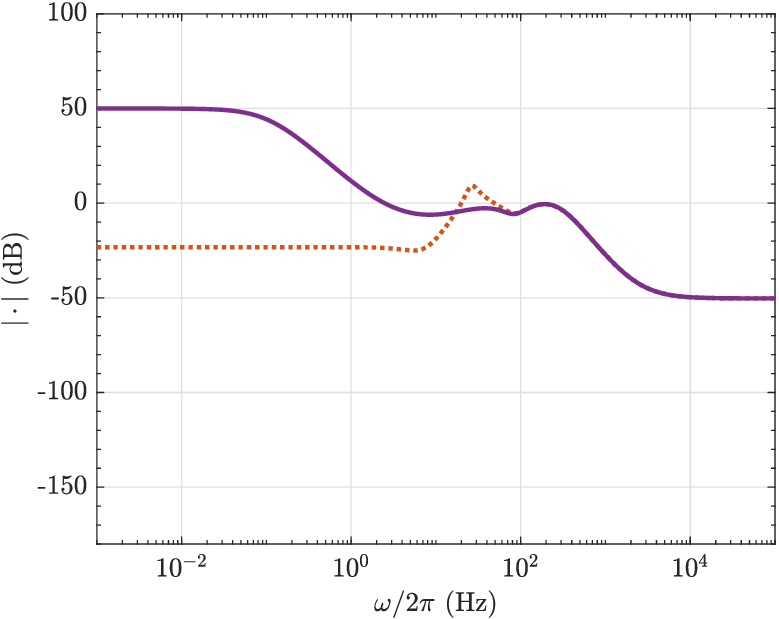}
\caption{Bode frequency responses for $\mathbb{K}\left(s\right)$ (${\color{mag1}\full}$) and $\mathbb{K}\left(s\right)\mathbb{S}_0\left(s\right)$ (${\color{red1}\dotted}$)}
\label{fig13}
\end{figure}

\begin{figure}
\centering
\includegraphics[width=0.4\linewidth]{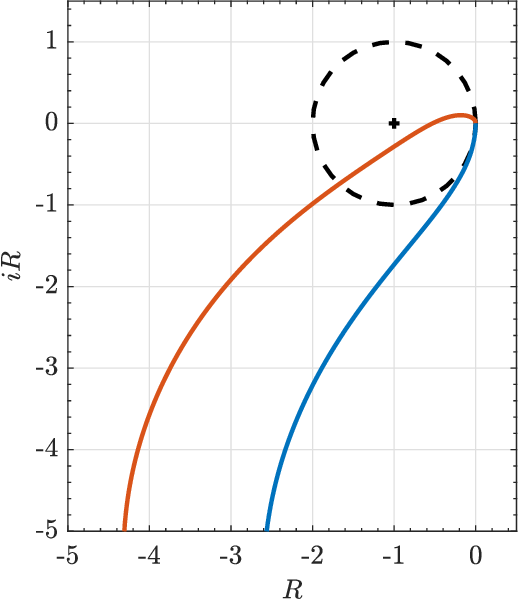}
\caption{Nyquist loci for $\mathbb{M}\left(s\right)$ (${\color{blue1}\full}$) and ${\mathbb{G}}_0\left(s\right)\mathbb{K}\left(s\right)$ (${\color{red1}\full}$)}
\label{fig08}
\end{figure}

\begin{figure}
\centering
\includegraphics[width=0.6\linewidth]{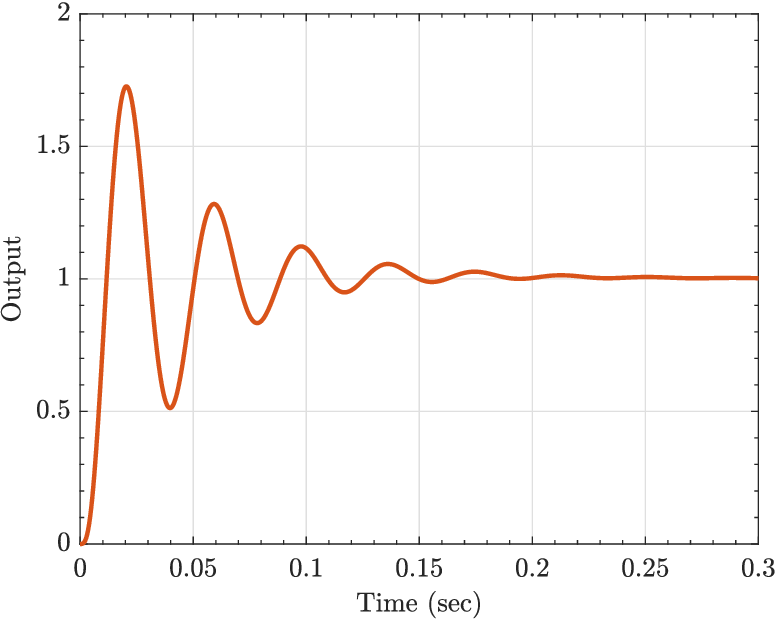}
\caption{Closed-loop unit step response}
\label{fig09}
\end{figure}

\begin{figure}
\centering
\begin{subfigure}[t]{\linewidth}
\centering\includegraphics[width=0.6\linewidth]{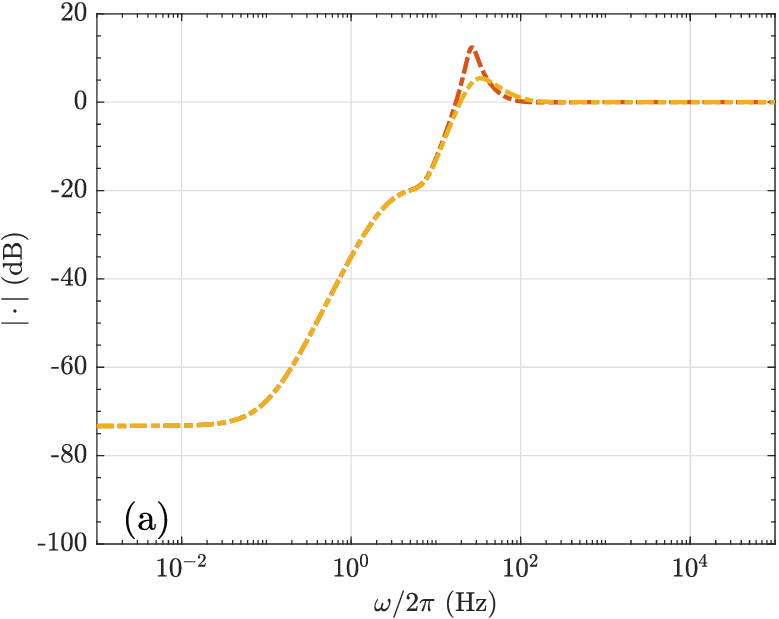}
\end{subfigure}
\medskip
\begin{subfigure}[t]{\linewidth}
\centering\includegraphics[width=0.6\linewidth]{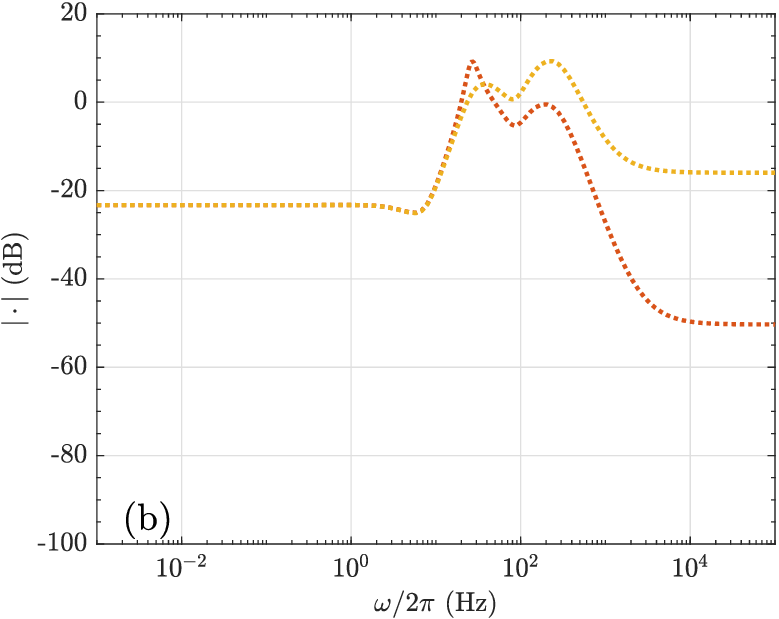}
\end{subfigure}
\caption{Bode frequency responses for (a) original (${\color{red1}\chain}$) and modified (${\color{yellow1}\chain}$) $\mathbb{S}_0\left(s\right)$; (b) original (${\color{red1}\dotted}$) and modified (${\color{yellow1}\dotted}$) $\mathbb{K}\left(s\right)\mathbb{S}_0\left(s\right)$}\label{fig10}
\end{figure}

\begin{figure}
\centering
\includegraphics[width=0.6\linewidth]{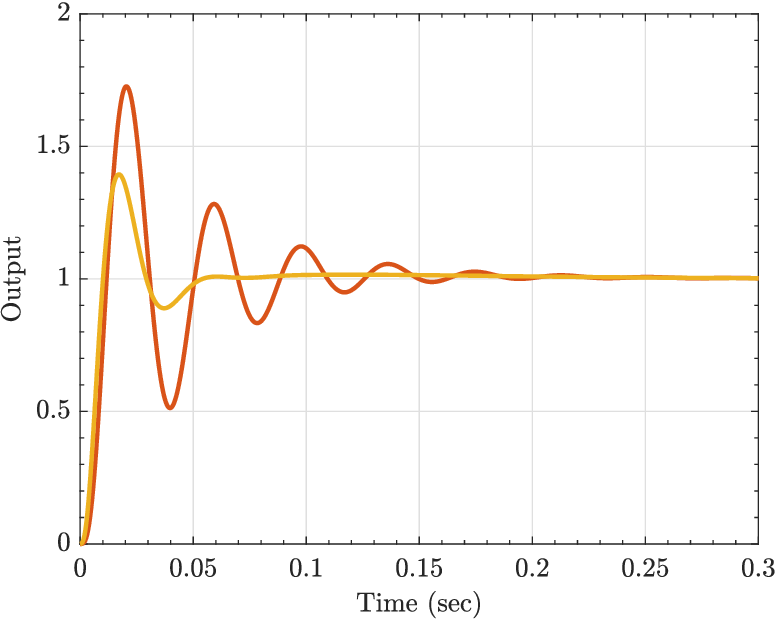}
\caption{Original (${\color{red1}\full}$) and modified (${\color{yellow1}\full}$) closed-loop unit step responses}
\label{fig11}
\end{figure}

\begin{figure}
\centering
\includegraphics[width=0.4\linewidth]{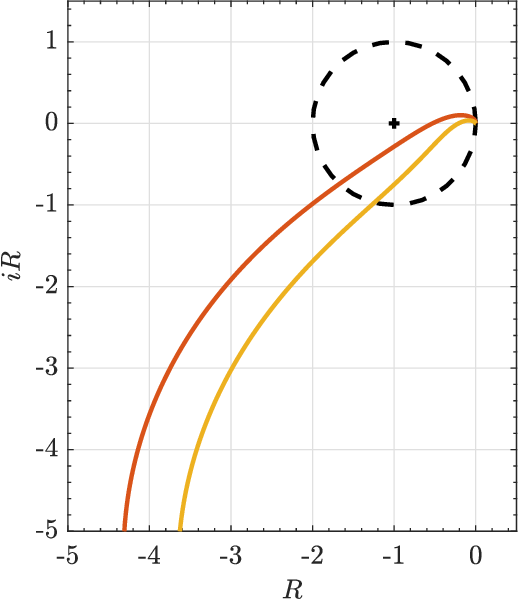}
\caption{Original (${\color{red1}\full}$) and modified (${\color{yellow1}\full}$) Nyquist loci for ${\mathbb{G}}_0\left(s\right)\mathbb{K}\left(s\right)$}
\label{fig12}
\end{figure}

We begin the design process with Fig.\ \ref{fig03}: Fig.\ \ref{fig03}(a) shows the frequency response for the nominal open-loop transfer function $\mathbb{M}_0\left(s\right)$ of the KBF error process whereas Fig.\ \ref{fig03}(b) shows the frequency response for the sensitivity function $(1+\mathbb{M}_0\left(s\right))^{-1}$. (Notice that $\mathbb{M}_0\left(s\right)$ has not been explicitly defined in the paper but is straightforward from the construction of $\mathbb{M}\left(s\right)$ in \eqref{eqMs}.) The low-frequency design specifications \eqref{eq_04} aim at providing good tracking and disturbance rejection in $\mathit{\Delta}^{\omega_0}$. The unity $\mathbb{M}_0$-gain crossover frequency $\omega_0$ is at about $40\times2\pi\,\text{rad}/\text{sec}$ and we let $\omega_{11}$ and $\omega_{12}$ be $1\times2\pi\,\text{rad}/\text{sec}$ and $7.5\times2\pi\,\text{rad}/\text{sec}$, respectively ($7.5\times2\pi\,\text{rad}/\text{sec}$ is approximately the resonant frequency of the plant). The remaining frequencies $\omega_{21}$ and $\omega_{22}$ for use in the high-frequency design specifications \eqref{eq_0s4} are, respectively, $120\times2\pi\,\text{rad}/\text{sec}$ and $450\times2\pi\,\text{rad}/\text{sec}$.

As mentioned in the last paragraph of the preceding section, the requirements in \eqref{aassqw} and \eqref{aasssqw} should be restricted to their boundaries so that the chances of ultimately finding an LQG compensator transfer function are improved. We obtain the following algebraic system of equations for $k=1,2$ and $m,p\ge 1$:
\begingroup
\allowdisplaybreaks
\begin{equation}
\left\{\begin{split}\label{algeq01}
\tau_{22}^p({\omega_{1k}^2+\tau_{21}^2})^{p/2}- \frac{1}{m_{1k}}\left|{{\mathbb{G}}_0\left(i\omega_{1k}\right)}\right|^{-1}({\omega_{1k}^2+\tau_{22}^2})^{p/2}&=0,\\
{\tau_{12}^m}({\omega_{2k}^2+\tau_{11}^2})^{m/2}-\frac{1}{m_{2k}}\left|\mathbb{M}\left(i\omega_{2k}\right)\right|\left|{\mathbb{G}}_0\left(i\omega_{2k}\right)\right|^{-1}{\tau_{11}^m}({\omega_{2k}^2+\tau_{12}^2 })^{m/2}&=0.
\end{split}\right.
\end{equation}
\endgroup
The next step therefore involves determining the bound pairs $m_{1k}$, $m_{2k}$ for which suitable solution pairs $\tau_{j1}$, $\tau_{j2}$ in the sense made precise in Section \ref{sec04} do (or do not) exist. It has been determined in preliminary studies that one would like to have (magnitudes $\left|\,\cdot\,\right|$ expressed in decibels, as is practical)
\begin{equation}\label{eq112ssa}
\left|\mathbb{S}\left(i\omega_{11}\right)\right|= -35\,\text{dB},\quad \left|\mathbb{S}\left(i\omega_{12}\right)\right|= -18\,\text{dB}
\end{equation} 
and
\begin{equation}\label{eq112ssb}
\left|\mathbb{K}\left(i\omega_{21}\right)\mathbb{S}\left(i\omega_{21}\right)\right|= -3\,\text{dB},\quad \left|\mathbb{K}\left(i\omega_{22}\right)\mathbb{S}\left(i\omega_{22}\right)\right|= -10\,\text{dB}.
\end{equation}
The unknowns $\tau_{j1}$, $\tau_{j2}$ are sought numerically for values of the $m_{1k}$, $m_{2k}$ in intervals around these points. One may choose to divide each of the four intervals either into an equal or unequal number of points according to the demands of computer time and required combinations. For this example we choose $7$ points in each of the intervals, which yields a total of $7^4= 2401$ possible combinations, and assume for the intervals
\begin{equation}\label{eq121112}
-35\,\text{dB}\le m_{11} \le -21\,\text{dB},\quad -18\,\text{dB}\le m_{12} \le -9\,\text{dB}
\end{equation}
and
\begin{equation}\label{eq12111s2}
-3\,\text{dB}\le m_{21} \le 11\,\text{dB},\quad -10\,\text{dB}\le m_{22} \le 3\,\text{dB}.
\end{equation}
Now standard solution algorithms of Matlab or Maple can be implemented to obtain numerical results, and at this point we should decide on the orders $m$ and $p$. Three cases are considered: $m=p=1$, \textbf{Case 1}, $m=p=2$, \textbf{Case 2}, and $m=3$, $p=2$, \textbf{Case 3}. These cases are not an arbitrary choice but serve to illustrate the radically different closed-loop behaviour we can obtain when weighting augmentation considerations are included in the LQG/LTR design problem formulation. (It is important to stress that it is difficult to predict a priori what order of weightings will be necessary to provide simultaneous adherence to the requirements in \eqref{aassqw} and \eqref{aasssqw}.)

For each case the results were computed iteratively with the use of a symbolic computation routine. Each iteration in the solution of \eqref{algeq01} corresponded to a unique combination of varying bound pairs $m_{11}$, $m_{12}$ and $m_{21}$, $m_{22}$ in the regions \eqref{eq121112} and \eqref{eq12111s2}, respectively. Numerical results for each of the combinations were obtained, checking throughout that the conditions in \eqref{eqass01sx} were satisfied. Whenever the conditions were satisfied, the solution pairs $\tau_{11}$, $\tau_{12}$ and $\tau_{21}$, $\tau_{22}$ were accepted as suitable and were subsequently saved for post-processing. The iterative solution routine is a very natural one and has the advantage that it is performed only once and is decoupled from the attainability problem (mentioned at the end of Section \ref{sec04}). This separation clearly simplifies the design process.

It is perhaps not too surprising that for \textbf{Case 1} no combinations could be found in the two regions \eqref{eq121112} and \eqref{eq12111s2} for which suitable solution pairs existed. Results for \textbf{Case 2} are presented in Figs.\ \ref{fig04b} and \ref{fig04bb} in the appendix. Suitable solution pairs were found for various combinations of bound pairs; however, we were not able to satisfy the latter requirement in \eqref{eq112ssb} (wherein the bound $m_{22}$ has the value $-10\,\text{dB}$). It was found by careful inspection of the results in Figs.\ \ref{fig04b} and \ref{fig04bb} that for $m_{11}=-35\,\text{dB}$, $m_{12}=-18\,\text{dB}$ and $m_{21}=-3\,\text{dB}$, $m_{22}=-1.333\,\text{dB}$ we have $\tau_{11}=330.09965$, $\tau_{12}=1.19229\times10^4$ and $\tau_{21}=17.75991$, $\tau_{22}=0.65409$ and it was seen that these are the best possible suitable solution pairs in this case. Results for \textbf{Case 3} are presented in Figs.\ \ref{fig05b} and \ref{fig05bb} in the appendix. As can be seen from the figures, in this case more combinations of bound pairs could be found for which suitable solution pairs existed, and we were finally able to satisfy simultaneously all the requirements in \eqref{eq112ssa} and \eqref{eq112ssb}. In a similar way to \textbf{Case 2}, we found that for $m_{11}=-35\,\text{dB}$, $m_{12}=-18\,\text{dB}$ and $m_{21}=-3\,\text{dB}$, $m_{22}=-10\,\text{dB}$ we had $\tau_{11}=488.55273$, $\tau_{12}=1.41057\times10^4$ and $\tau_{21}=17.75991$, $\tau_{22}=0.65409$. (Note that the values of $m_{11}$ and $m_{12}$ are the same as in \textbf{Case 2}, thus the computation yields the same values for $\tau_{21}$ and $\tau_{22}$ found in \textbf{Case 2} and so there is a suitable solution pair $\tau_{21}$, $\tau_{22}$ in \textbf{Case 3} which is the same as in \textbf{Case 2}.)

We continue to consider \textbf{Case 3}. The above results give for the two weightings the following transfer functions:
\begin{equation*}
\mathcal{W}_1\left(s\right)=\frac{2.40685\times10^4(s+488.55273)^3}{(s+1.41057\times10^4)^3},\quad \mathcal{W}_2\left(s\right)=\frac{0.42783(s+17.75991)^2}{(s+0.65409)^2},
\end{equation*}
and the resulting LQG compensator transfer function is
\begin{equation*}
\mathbb{K}\left(s\right)=\frac{\begin{gathered}-1.51903\times10^8s^8 -6.50277\times10^{12}s^7 -9.38834\times10^{16}s^6-4.73061 \times10^{20}s^5\\
-2.40161\times10^{23}s^4-1.46194\times10^{26}s^3 -1.49704\times10^{28}s^2\\
-4.40097 \times10^{29}s-3.50558 \times10^{30}\end{gathered}}{\begin{gathered}s^9+4.97905\times10^{10}s^8+2.65614\times10^{14}s^7+6.44707\times10^{17}s^6+8.45845\times10^{20}s^5\\
+5.83906\times10^{23}s^4+1.98384\times10^{26}s^3+2.62420\times10^{28}s^2\\
+3.40755\times10^{28}s+1.11166\times10^{28}\end{gathered}}.
\end{equation*}
In Fig.\ \ref{fig04} we compare the relevant frequency responses associated with the KBF with those already shown in Fig.\ \ref{fig03}. The effect of the weighting $\mathcal{W}_2\left(s\right)$ is illustrated in Fig.\ \ref{fig04}(a), and Fig.\ \ref{fig04}(b) compares the frequency responses for $(1+\mathbb{M}_0\left(s\right))^{-1}$ and $(1+\mathbb{M}\left(s\right))^{-1}$. Results for the LQG open and closed loops are shown in Fig.\ \ref{fig05}. Fig.\ \ref{fig05}(a) compares the frequency response for the open-loop transfer function $\mathbb{M}\left(s\right)$ of the KBF error process with the frequency response for the LQG open-loop transfer function ${\mathbb{G}}_0\left(s\right)\mathbb{K}\left(s\right)$, and shows the effect of the weighting $\mathcal{W}_1\left(s\right)$. The corresponding (nominal) closed-loop frequency responses are shown in Fig.\ \ref{fig05}(b) and Fig.\ \ref{fig13}. Also shown in Fig.\ \ref{fig13} is the frequency response for the LQG compensator transfer function $\mathbb{K}\left(s\right)$. The results are summarised in Tab.\ \ref{tab1}. We conclude that all requirements are met very well.

\begin{table}
\centering
   \begin{tabular}{ccrrcrr}
   \toprule\toprule
   $k$ & $\omega_{1k}/\left(2\pi\,\text{Hz}\right)$ & $m_{1k}/\text{dB}$ & $\left|\mathbb{S}_0\left(i\omega_{1k}\right)\right|/\text{dB}$ & $\omega_{2k}/\left(2\pi\,\text{Hz}\right)$ &  $m_{2k}/\text{dB}$ & $\left|\mathbb{K}\left(i\omega_{2k}\right)\mathbb{S}_0\left(i\omega_{2k}\right)\right|/\text{dB}$  \\ 
   \midrule
   $1$ & $1$ & $-35$ & $-34.99498$ & $120$ & $-3$ & $-2.94387$ \\
   $2$ & $7.5$ & $-18$ & $-17.64842$ & $450$ & $-10$ & $-10.00115$  \\
   \bottomrule
   \end{tabular}
   \caption{Results for \textbf{Case 3}}\label{tab1}
\end{table}

Let us now make some comments about the behaviour of the closed loop in the crossover frequency range. The results indicate a peak of the frequency response for the sensitivity function $\mathbb{S}_0\left(s\right)$ in the crossover frequency range which does not occur in the frequency response for $(1+\mathbb{M}\left(s\right))^{-1}$, as seen in Fig.\ \ref{fig05}(b). This effect is also clear from the Nyquist plots in Fig.\ \ref{fig08} (as we have explained in the paragraph following \eqref{eqineqk1}). The problem here is not so much that peaking occurs (which is unavoidable) but that it possibly is ``too large'' and causes a ``too lightly damped'' oscillatory step response of the closed-loop system, as shown in the unit step response in Fig.\ \ref{fig09}. In particular, since large sensitivity magnitudes generally imply that the Nyquist locus for the LQG open-loop transfer function is rather close to the critical point, it follows that associated robustness margins will be too small (cf.\ Fig.\ \ref{fig08}). We may therefore wish to modify our LQG design by choosing larger values for the upper bounds on the controller noise sensitivity magnitudes of the LQG closed loop. For example, upon further detailed inspection of the results in Figs.\ \ref{fig05b} and \ref{fig05bb}, it was found that for $m_{21}=4\,\text{dB}$, $m_{22}=3\,\text{dB}$ we have $\tau_{11}=752.64610$, $\tau_{12}=5819.48938$. The closed-loop frequency responses are compared in Fig.\ \ref{fig10} for the original (i.e.\ with $\tau_{11}=488.55273$, $\tau_{12}=1.41057\times10^4$) and modified designs. The corresponding unit step response results are shown in Fig.\ \ref{fig11}, and in Fig.\ \ref{fig12} the Nyquist loci are compared. As expected, the robustness margin increases with deceasing $\mathbb{S}_0$-gain. The gain and phase margins are found to be approximately $4.8\,\text{dB}$ and $35.2^\circ$, respectively, as compared to approximately only $1.9\,\text{dB}$ and $15.1^\circ$ for the original design.

\section{Conclusion}

The approach presented in this expository paper provides a straightforward and systematic treatment of the LQG/LTR design problem for finite-dimensional SISO control systems based on considerations of weighting augmentation. An interesting feature of this approach is that it has incorporated aspects of classical loop shaping and that no complicated control optimisation has been carried out, i.e.\ the LQG compensator transfer function is found using a completely algebraic procedure.

We decided not to allow non-minimum-phase plants or plants with time delays in the current approach, an assumption which is well suited to the expository spirit of the paper. However, such system properties, individually or combined, are of some practical importance and therefore are a natural future extension. Another future extension of the approach should be concerned with the attainability problem and understanding the limitations as far as the closed-loop behaviour in the crossover frequency range is concerned. From a technical point of view, a major challenge concerns solving the algebraic system of equations \eqref{algeq01}, specifically seeking conditions under which solution pairs which are suitable exist. The numerical aspects of finding such solution pairs are especially challenging when one deals with inequality constraints and weightings of higher order. For example, our method requires the solution of a system of four non-linear algebraic equations, and, for weightings of higher order, the method becomes computationally cumbersome. This is yet another subject for future study, to which we hope to return in a later paper.

Finally, let us mention that the experimental verification of the LQG compensator is currently underway, the results of which will provide a much needed answer as to the practical applicability of our approach to the LQG/LTR design problem.

\bigskip\noindent
\textbf{Acknowledgments.} This paper was written while the first author was visiting the Antennas \& Electromagnetics Research Group, Queen Mary University of London. The support of the institution is greatly appreciated.

\bibliographystyle{plain}
\bibliography{BibLio01}

\newpage

\appendix

\section*{Appendix}
\renewcommand{\thefigure}{A\arabic{figure}}

\setcounter{figure}{0}

\begin{figure}[hbt!]
\centering
\begin{subfigure}[t]{\linewidth}
\centering\includegraphics[width=0.84\linewidth]{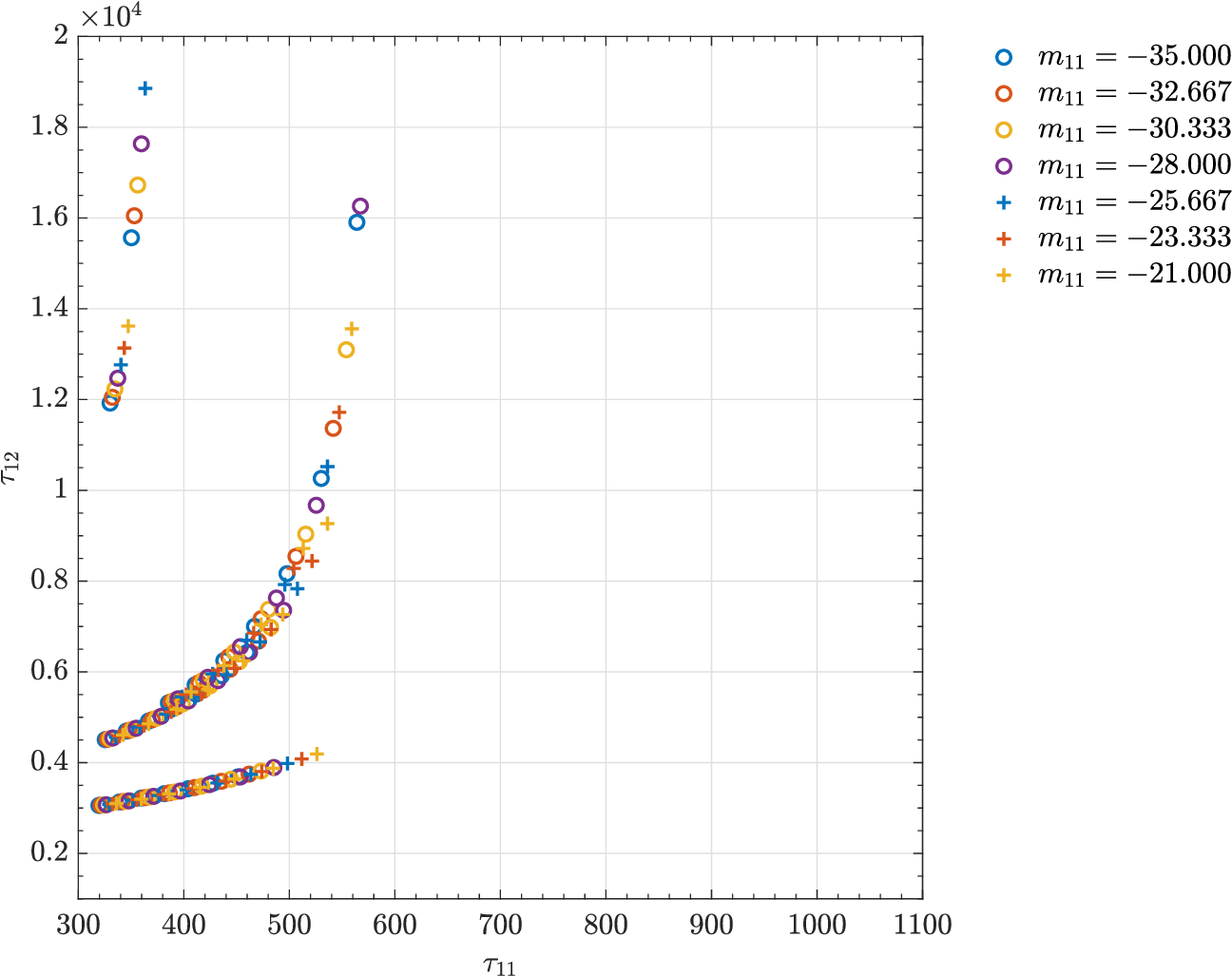}
\end{subfigure}
\medskip
\begin{subfigure}[t]{\linewidth}
\centering\includegraphics[width=0.84\linewidth]{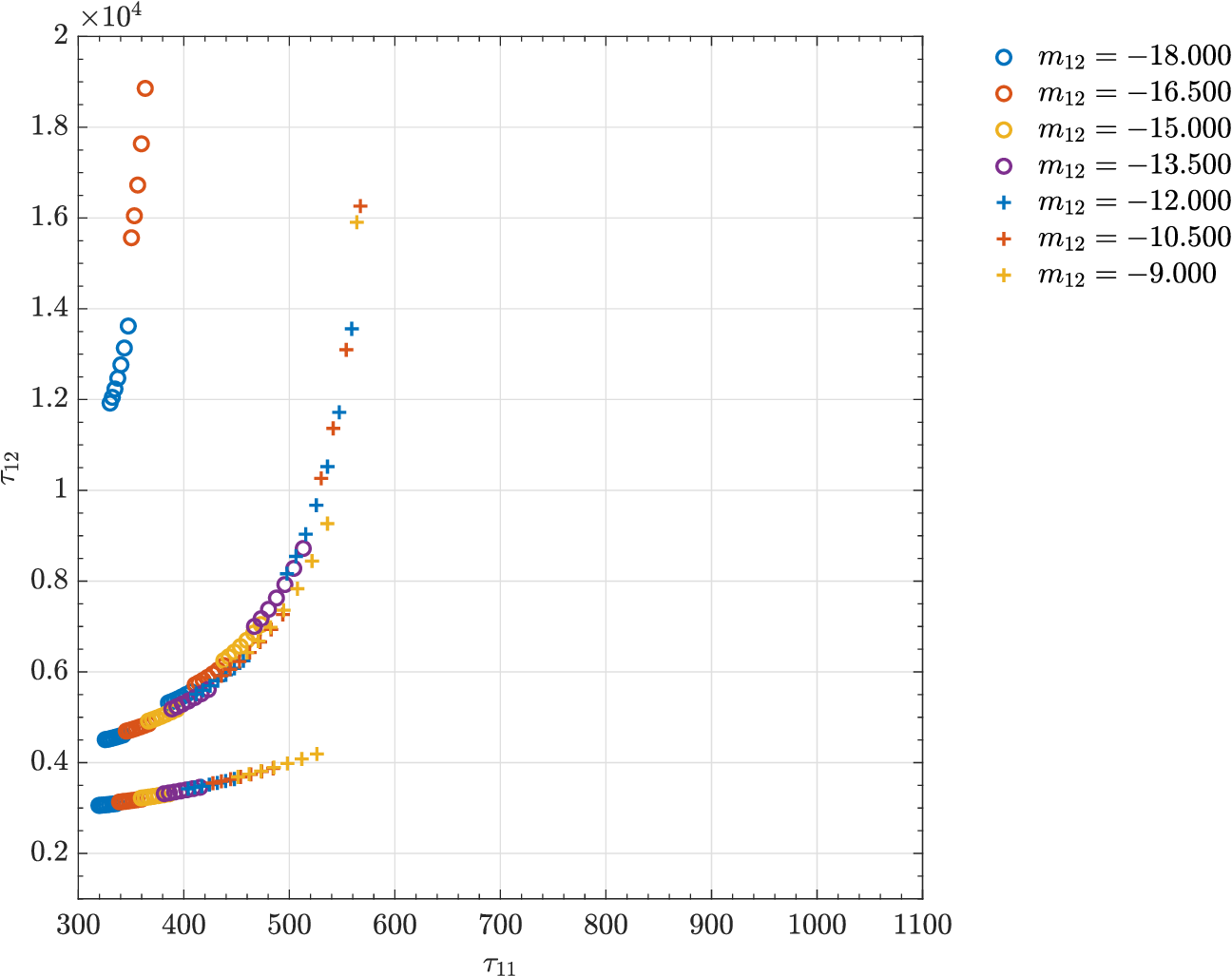}
\end{subfigure}
\caption{Values for \textbf{Case 2} of suitable solution pairs $\tau_{11}$, $\tau_{12}$ for combinations of bound pairs $m_{11}$, $m_{12}$ and $m_{21}$, $m_{22}$}
\end{figure}
\begin{figure}[hbt!]
\setcounter{figure}{0}
\centering
\begin{subfigure}[t]{\linewidth}
\centering\includegraphics[width=0.84\linewidth]{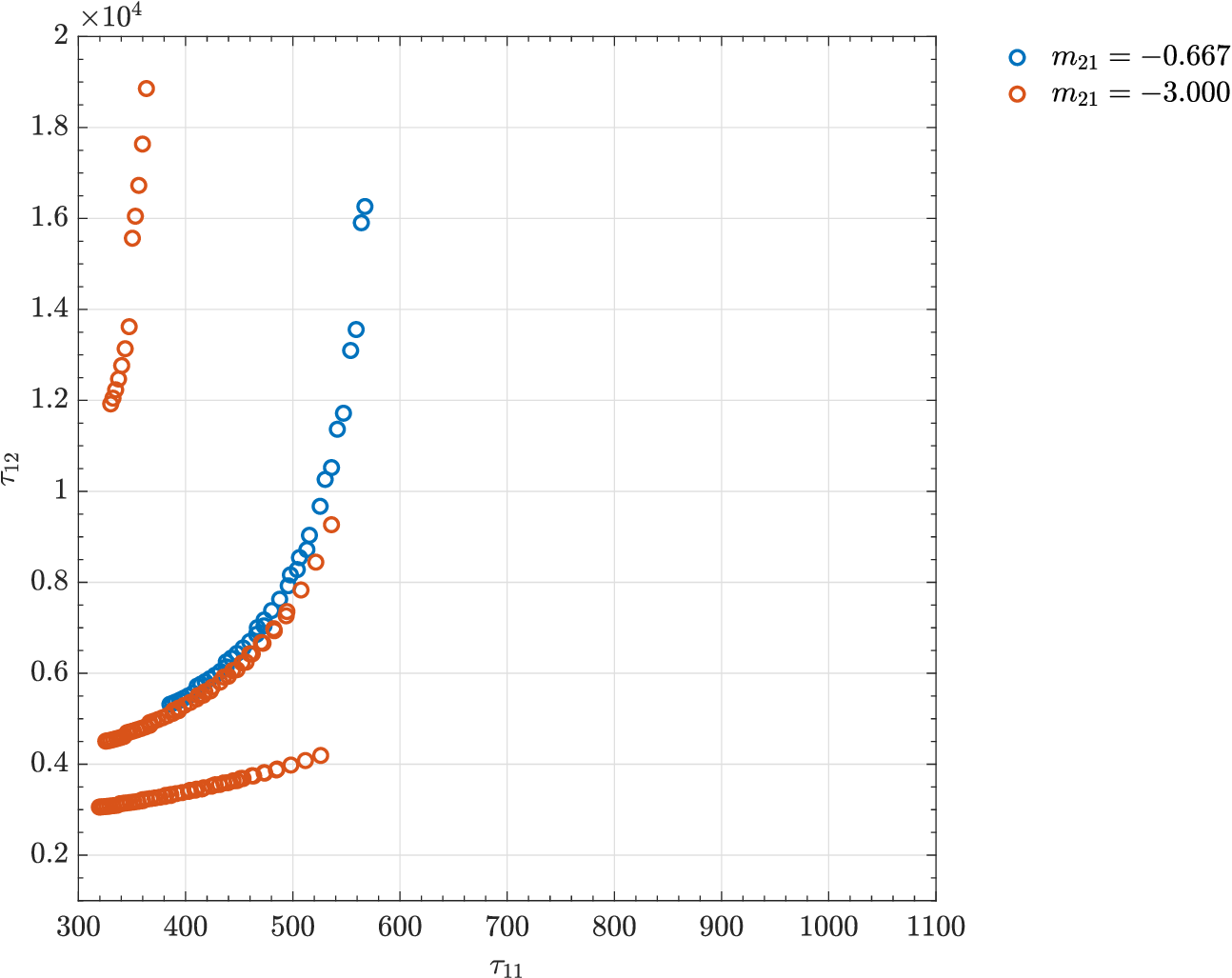}
\end{subfigure}
\medskip
\begin{subfigure}[t]{\linewidth}
\centering\includegraphics[width=0.84\linewidth]{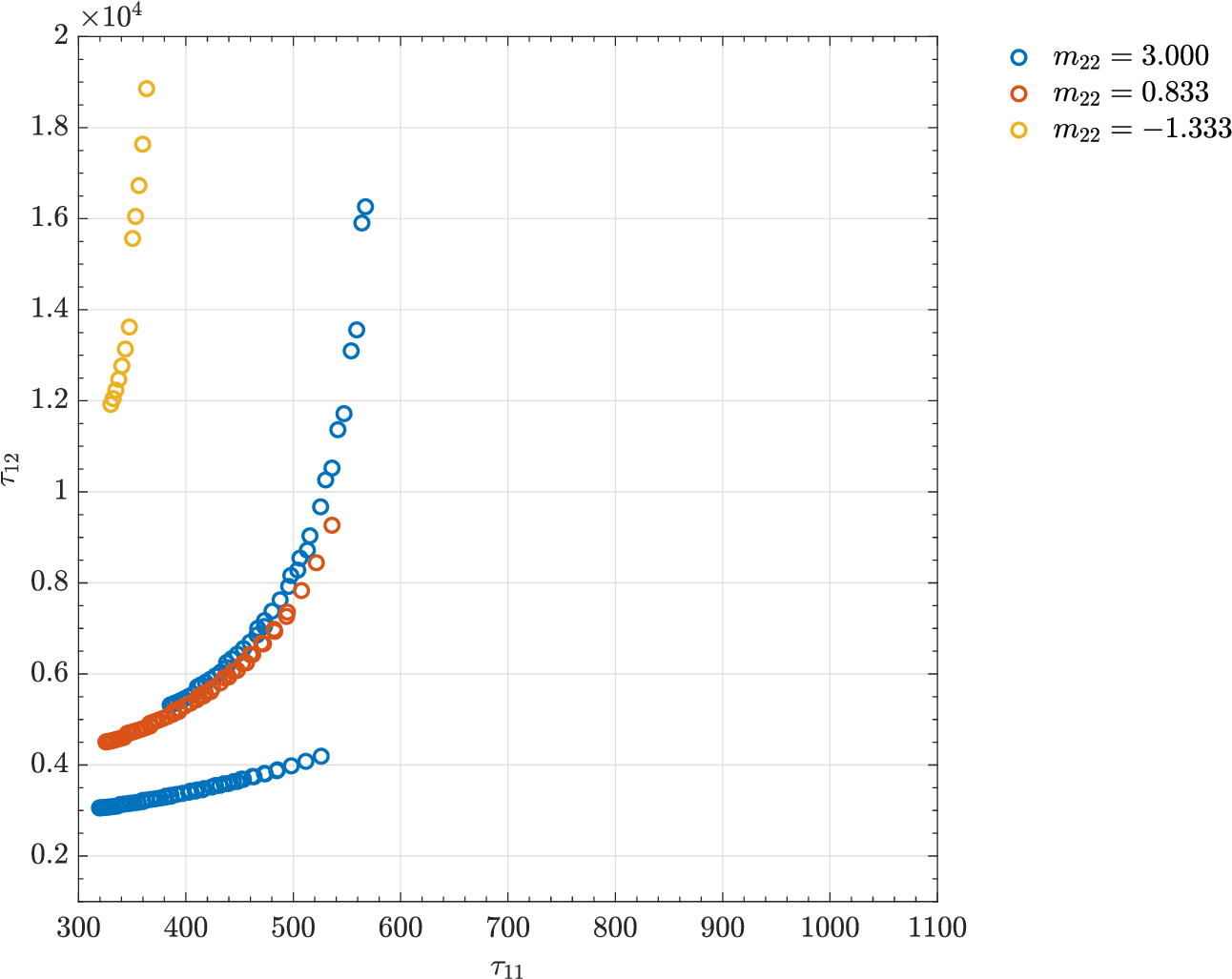}
\end{subfigure}
\caption{Values for \textbf{Case 2} of suitable solution pairs $\tau_{11}$, $\tau_{12}$ for combinations of bound pairs $m_{11}$, $m_{12}$ and $m_{21}$, $m_{22}$ (cont.)}\label{fig04b}
\end{figure}

\begin{figure}[hbt!]
\centering
\begin{subfigure}[t]{\linewidth}
\centering\includegraphics[width=0.84\linewidth]{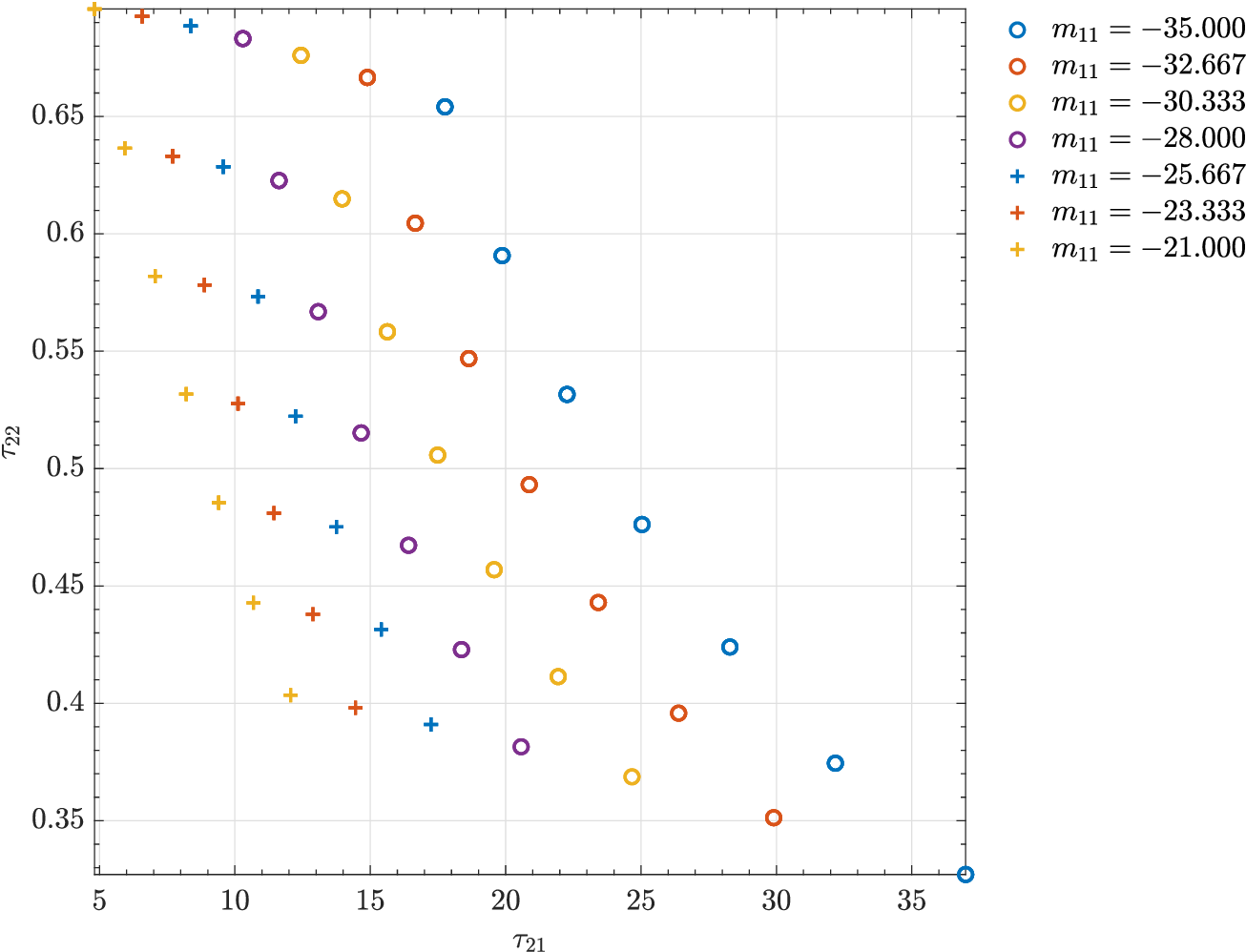}
\end{subfigure}
\medskip
\begin{subfigure}[t]{\linewidth}
\centering\includegraphics[width=0.84\linewidth]{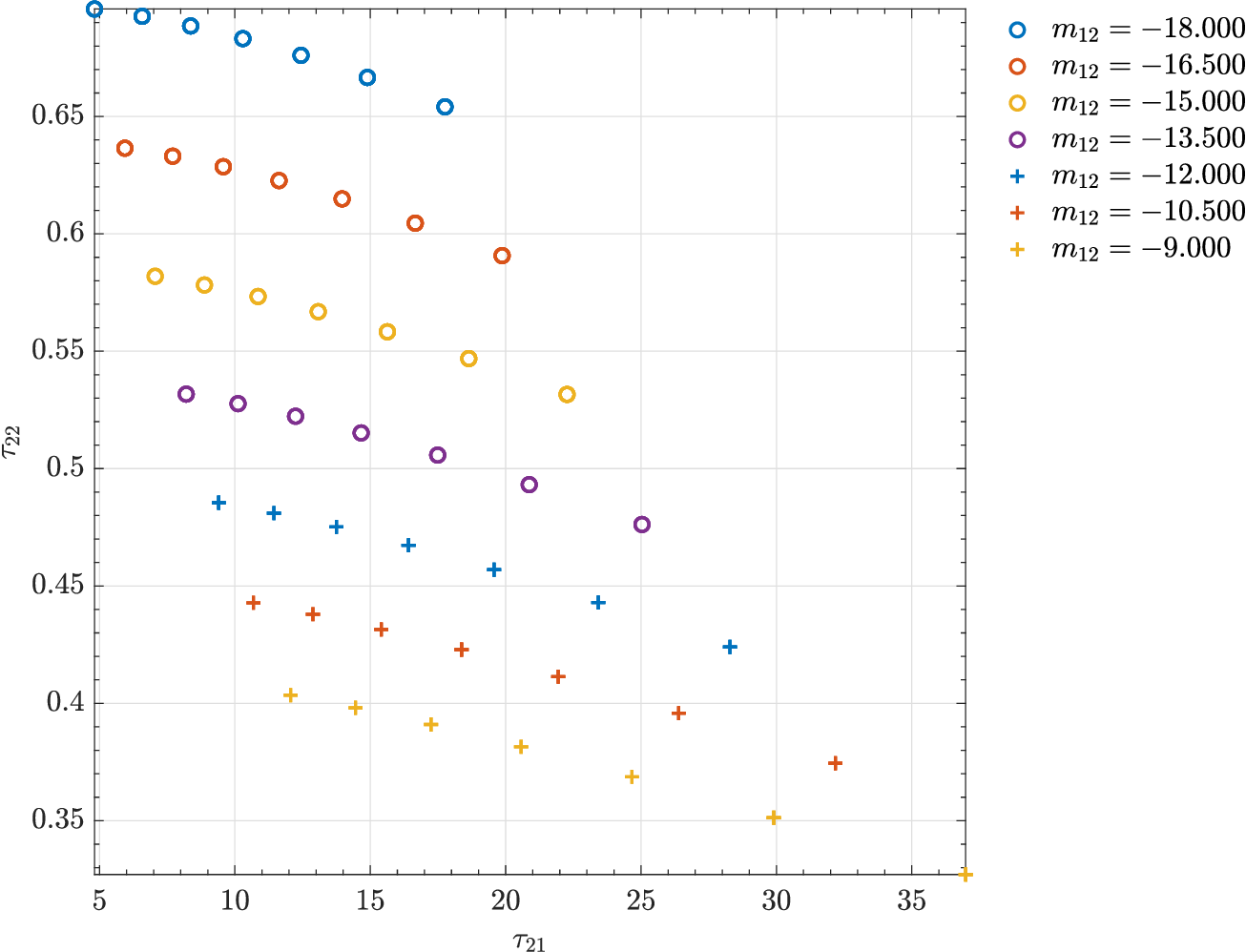}
\end{subfigure}
\caption{Values for \textbf{Case 2} of suitable solution pairs $\tau_{21}$, $\tau_{22}$ for combinations of bound pairs $m_{11}$, $m_{12}$ and $m_{21}$, $m_{22}$}
\end{figure}
\begin{figure}[hbt!]
\setcounter{figure}{1}
\centering
\begin{subfigure}[t]{\linewidth}
\centering\includegraphics[width=0.84\linewidth]{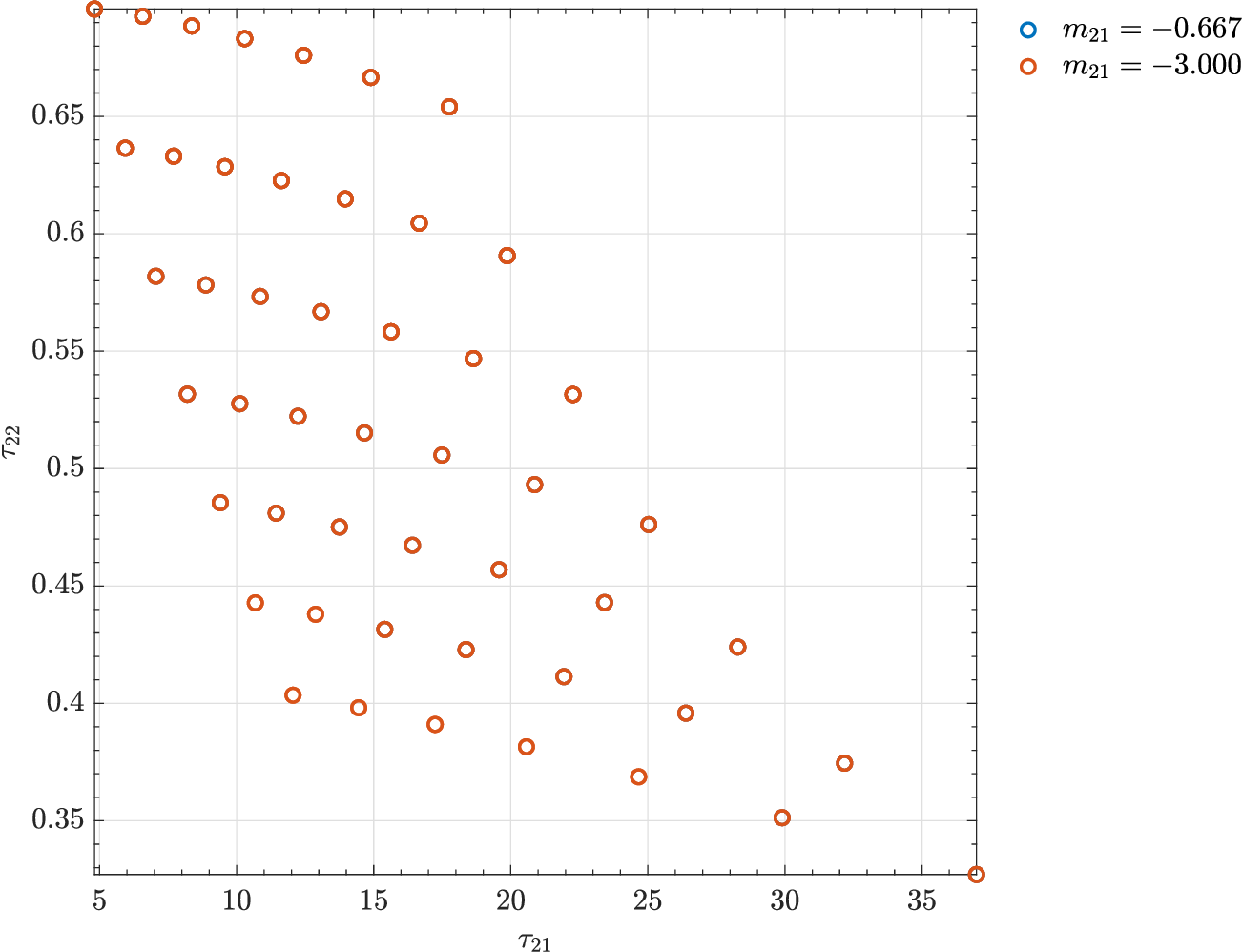}
\end{subfigure}
\medskip
\begin{subfigure}[t]{\linewidth}
\centering\includegraphics[width=0.84\linewidth]{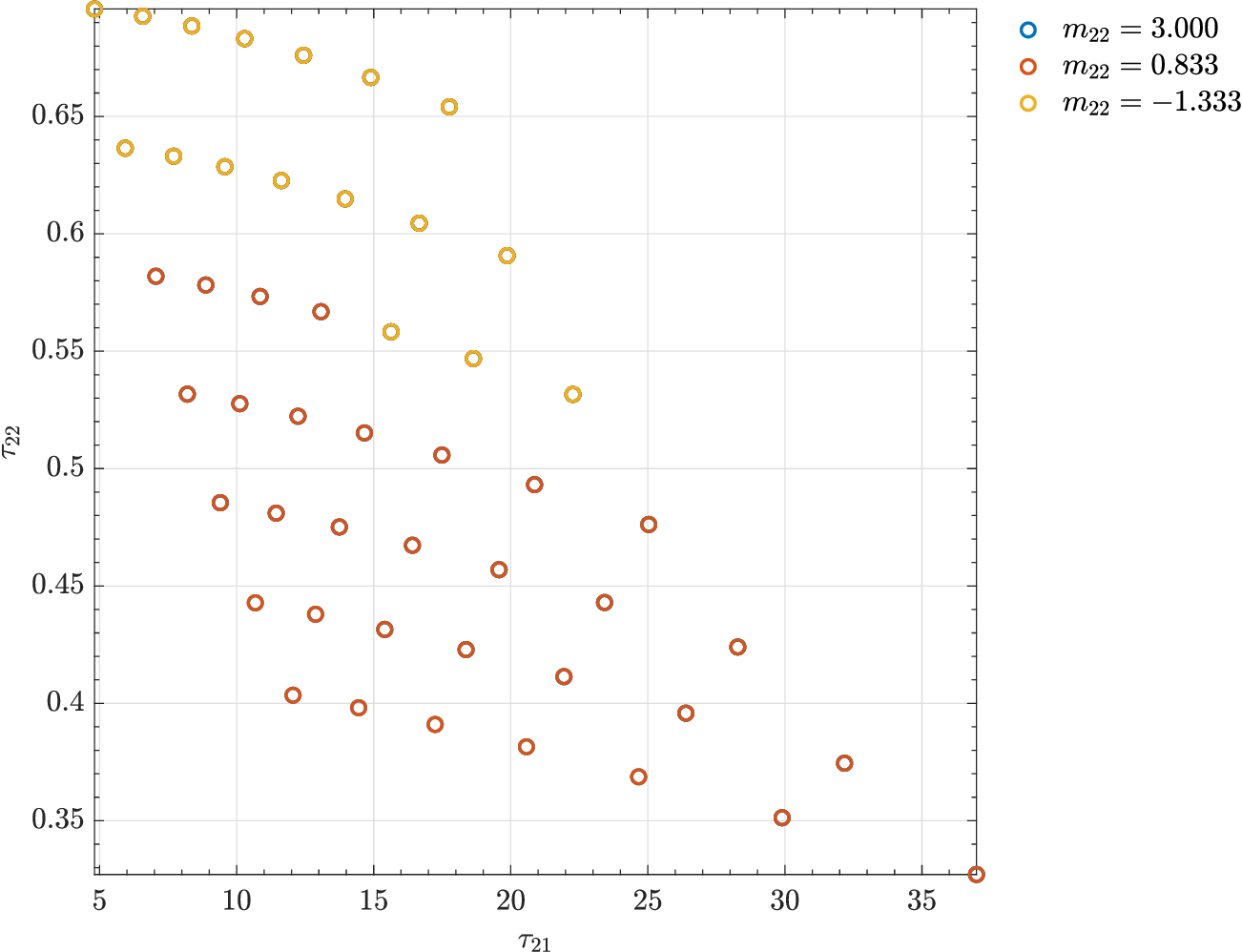}
\end{subfigure}
\caption{Values for \textbf{Case 2} of suitable solution pairs $\tau_{21}$, $\tau_{22}$ for combinations of bound pairs $m_{11}$, $m_{12}$ and $m_{21}$, $m_{22}$ (cont.)}\label{fig04bb}
\end{figure}

\begin{figure}[hbt!]
\centering
\begin{subfigure}[t]{\linewidth}
\centering\includegraphics[width=0.84\linewidth]{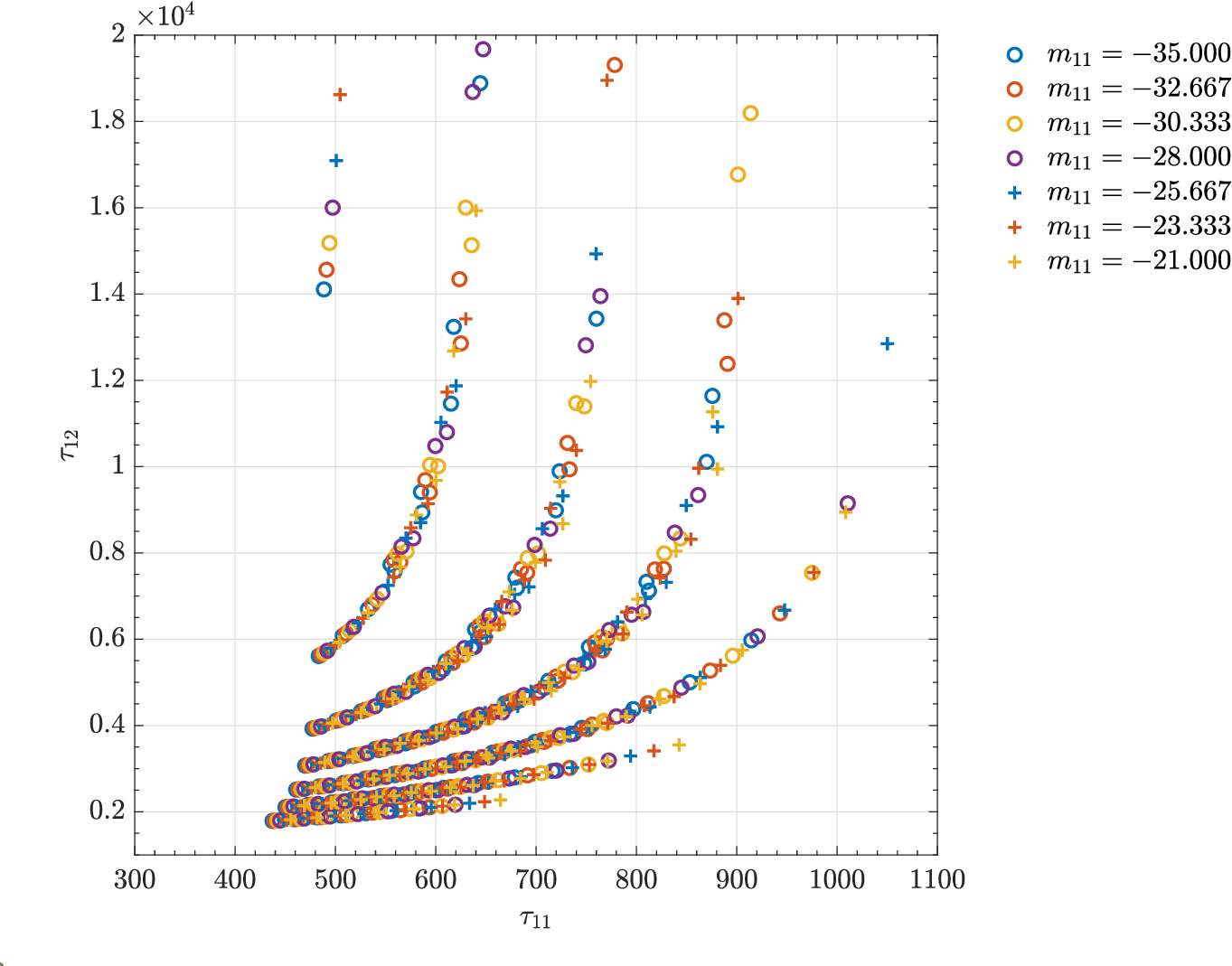}
\end{subfigure}
\medskip
\begin{subfigure}[t]{\linewidth}
\centering\includegraphics[width=0.84\linewidth]{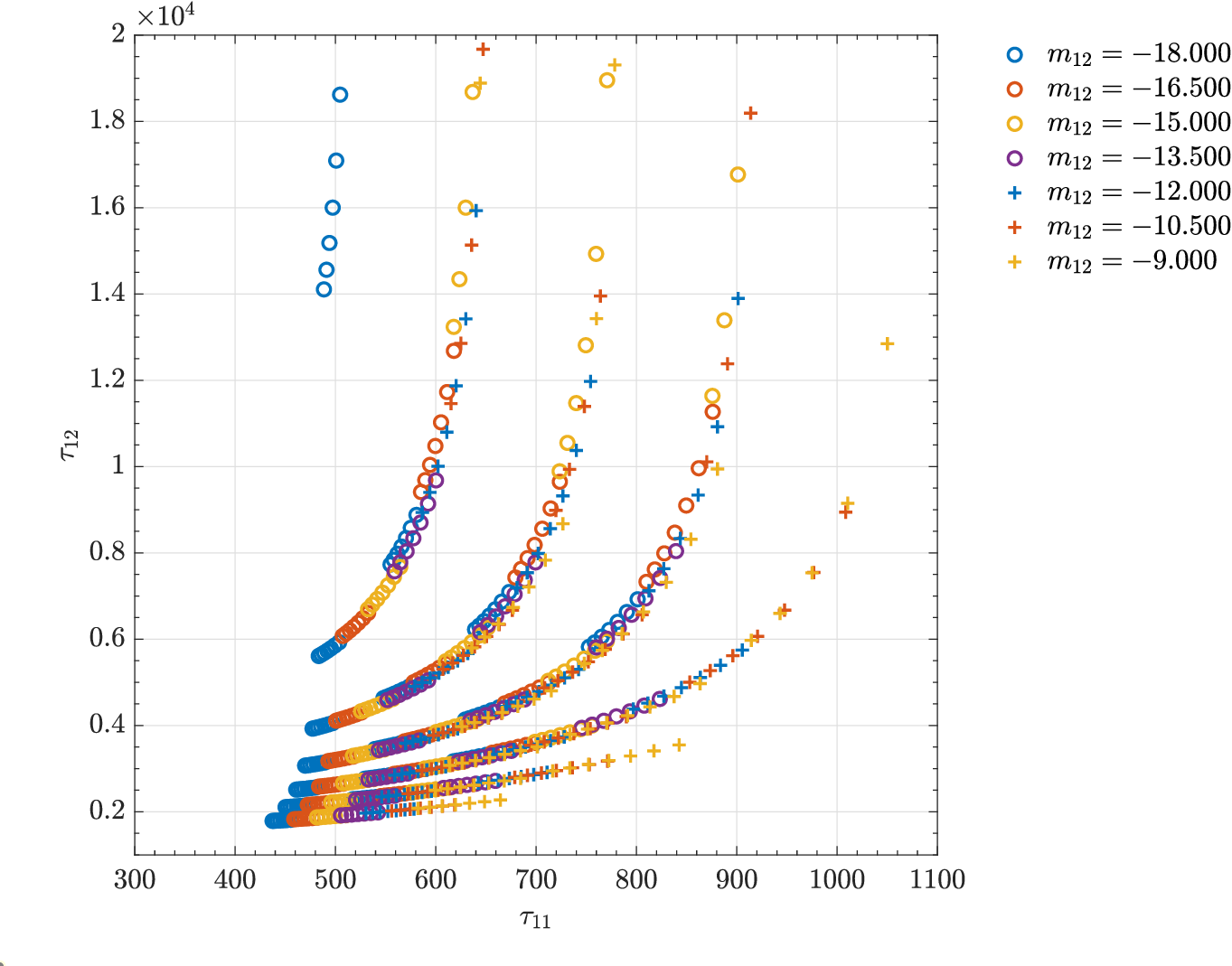}
\end{subfigure}
\caption{Values for \textbf{Case 3} of suitable solution pairs $\tau_{11}$, $\tau_{12}$ for combinations of bound pairs $m_{11}$, $m_{12}$ and $m_{21}$, $m_{22}$}
\end{figure}
\begin{figure}[hbt!]
\setcounter{figure}{2}
\centering
\begin{subfigure}[t]{\linewidth}
\centering\includegraphics[width=0.84\linewidth]{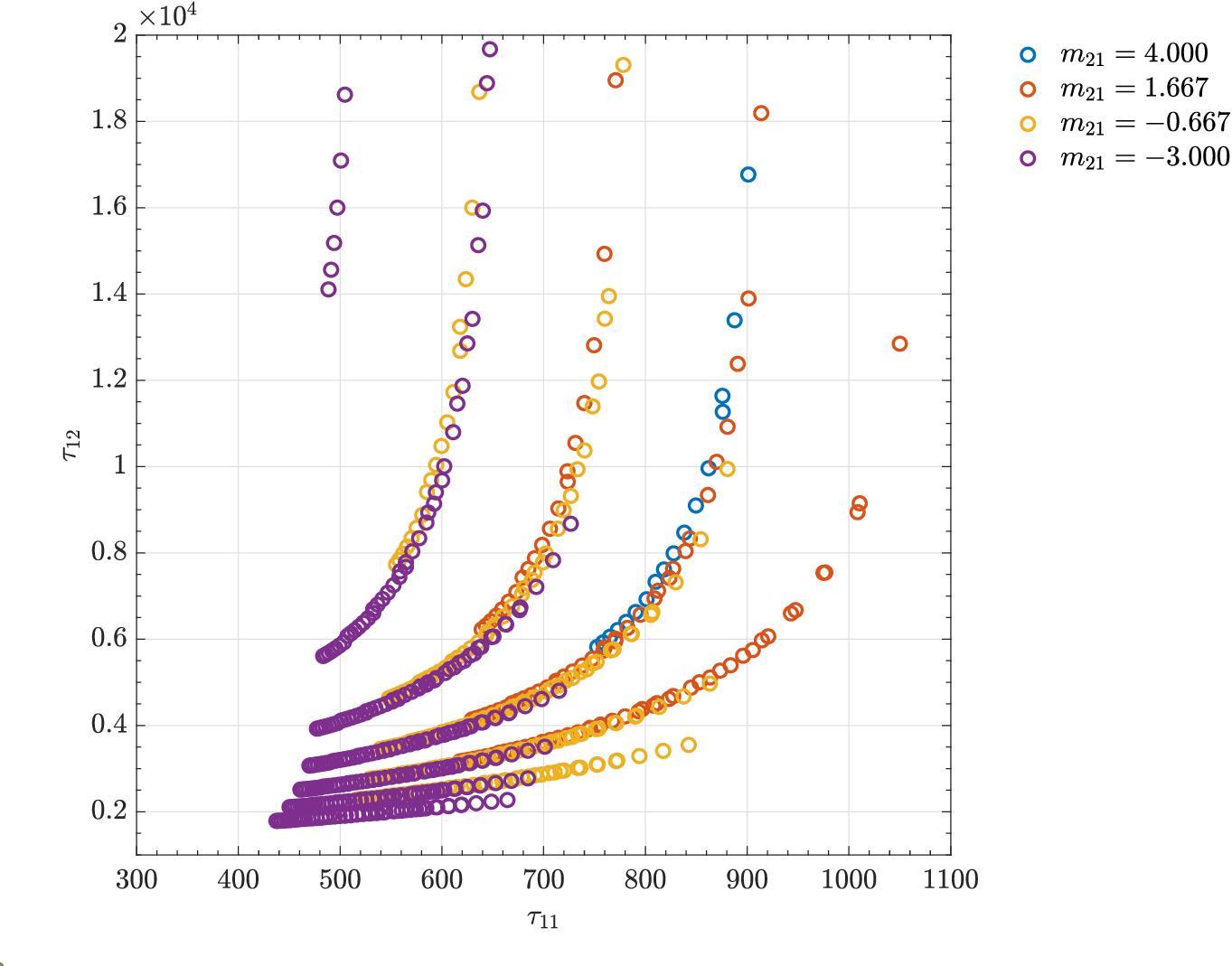}
\end{subfigure}
\medskip
\begin{subfigure}[t]{\linewidth}
\centering\includegraphics[width=0.84\linewidth]{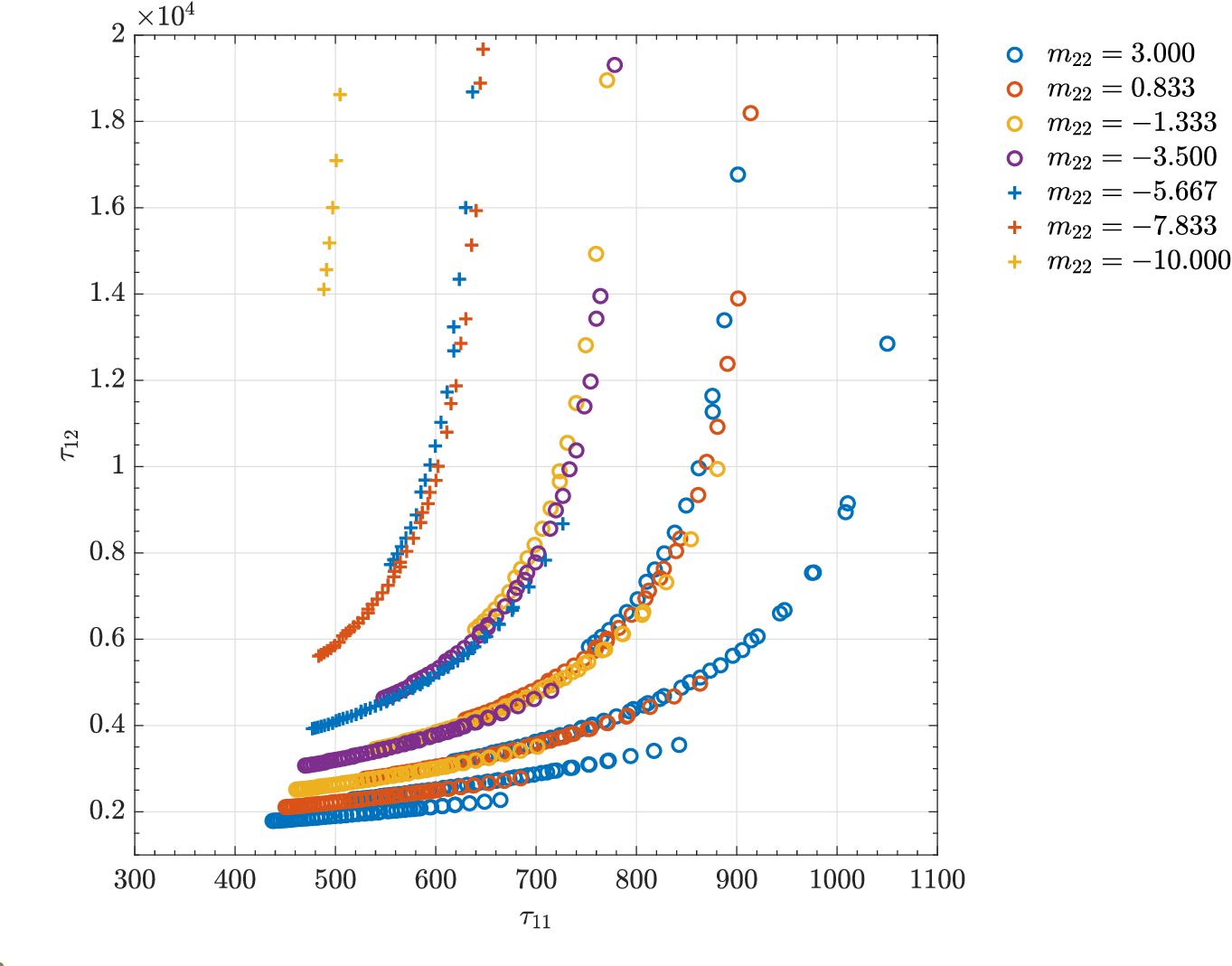}
\end{subfigure}
\caption{Values for \textbf{Case 3} of suitable solution pairs $\tau_{11}$, $\tau_{12}$ for combinations of bound pairs $m_{11}$, $m_{12}$ and $m_{21}$, $m_{22}$ (cont.)}\label{fig05b}
\end{figure}

\begin{figure}[hbt!]
\centering
\begin{subfigure}[t]{\linewidth}
\centering\includegraphics[width=0.84\linewidth]{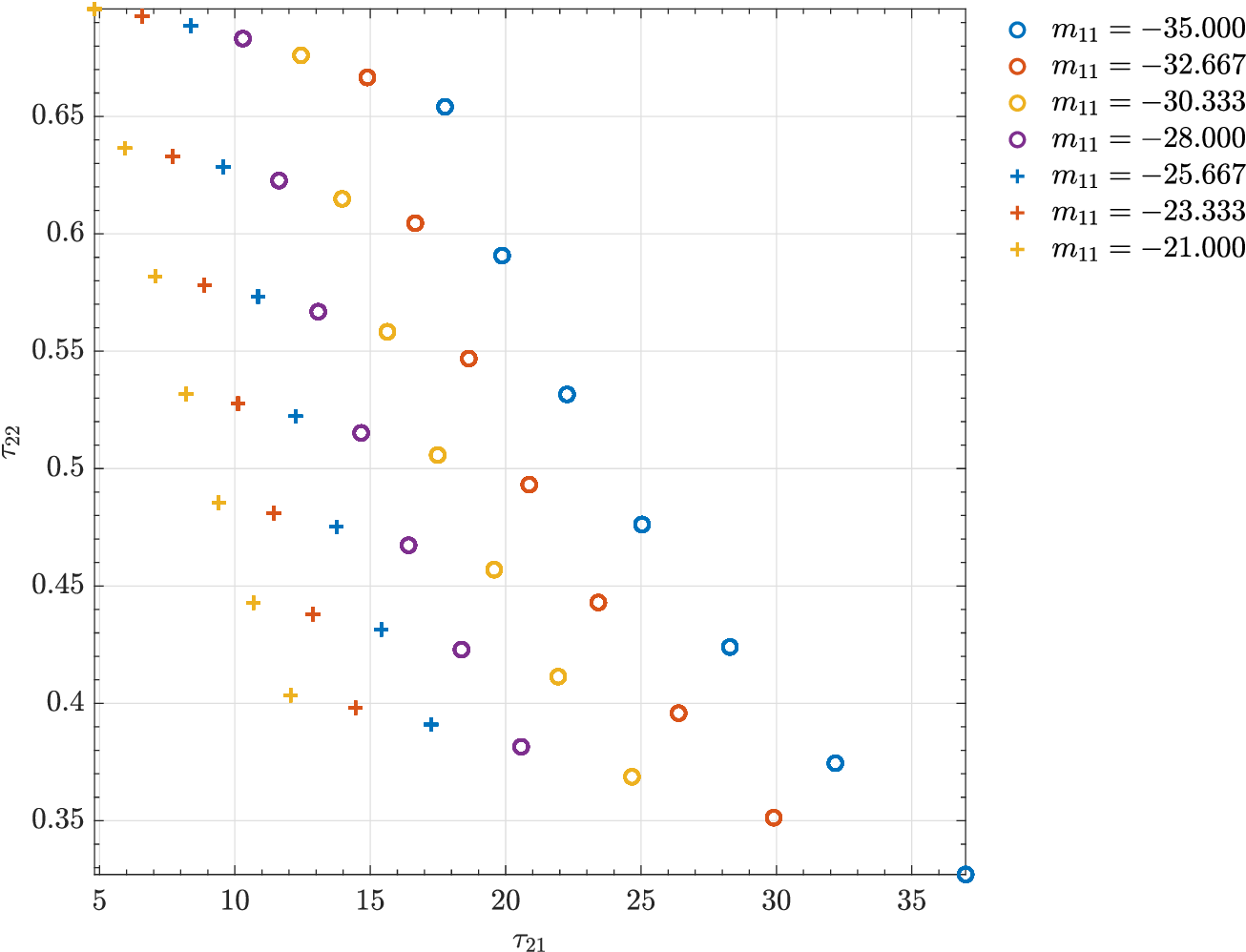}
\end{subfigure}
\medskip
\begin{subfigure}[t]{\linewidth}
\centering\includegraphics[width=0.84\linewidth]{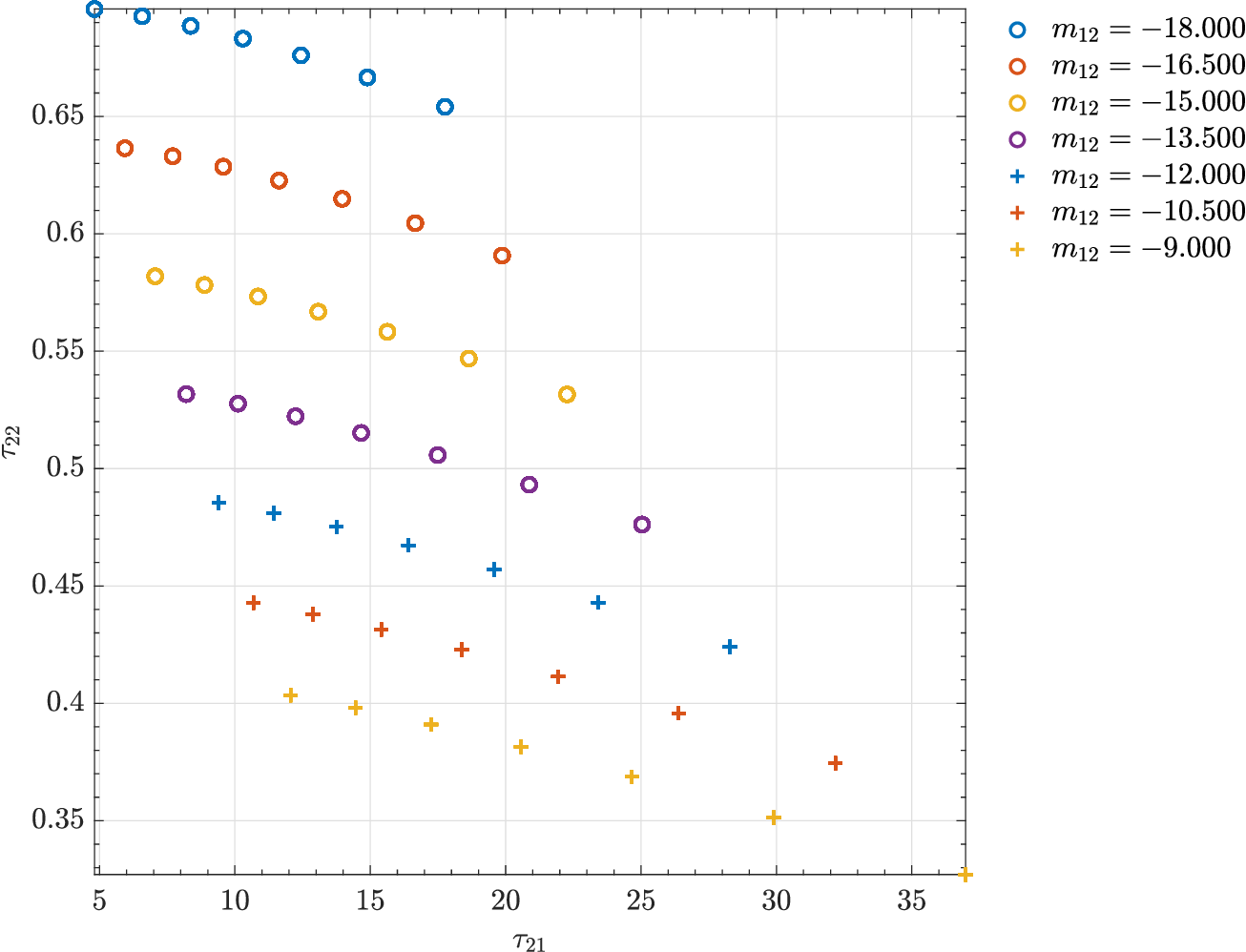}
\end{subfigure}
\caption{Values for \textbf{Case 3} of suitable solution pairs $\tau_{21}$, $\tau_{22}$ for combinations of bound pairs $m_{11}$, $m_{12}$ and $m_{21}$, $m_{22}$}
\end{figure}
\begin{figure}[hbt!]
\setcounter{figure}{3}
\centering
\begin{subfigure}[t]{\linewidth}
\centering\includegraphics[width=0.84\linewidth]{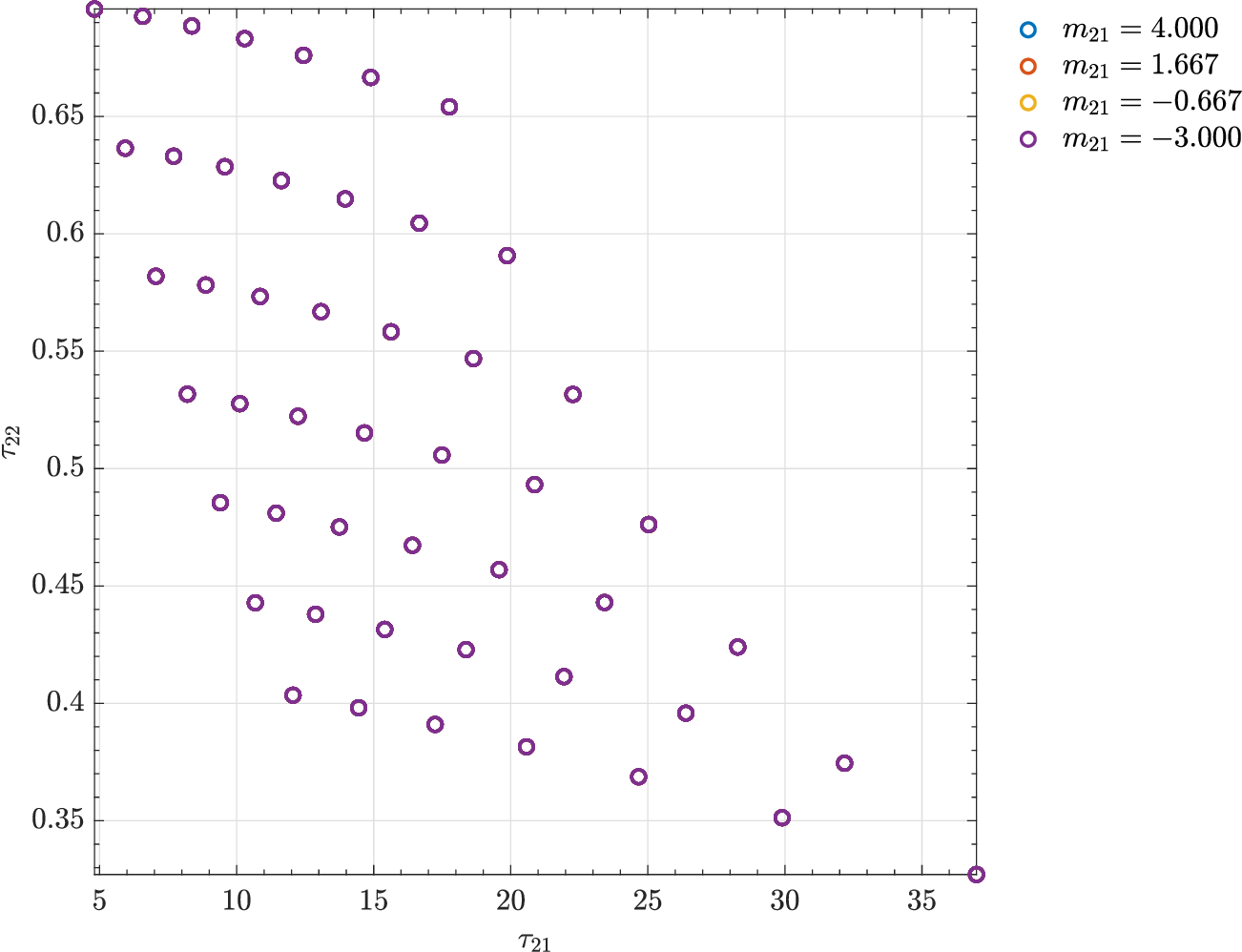}
\end{subfigure}
\medskip
\begin{subfigure}[t]{\linewidth}
\centering\includegraphics[width=0.84\linewidth]{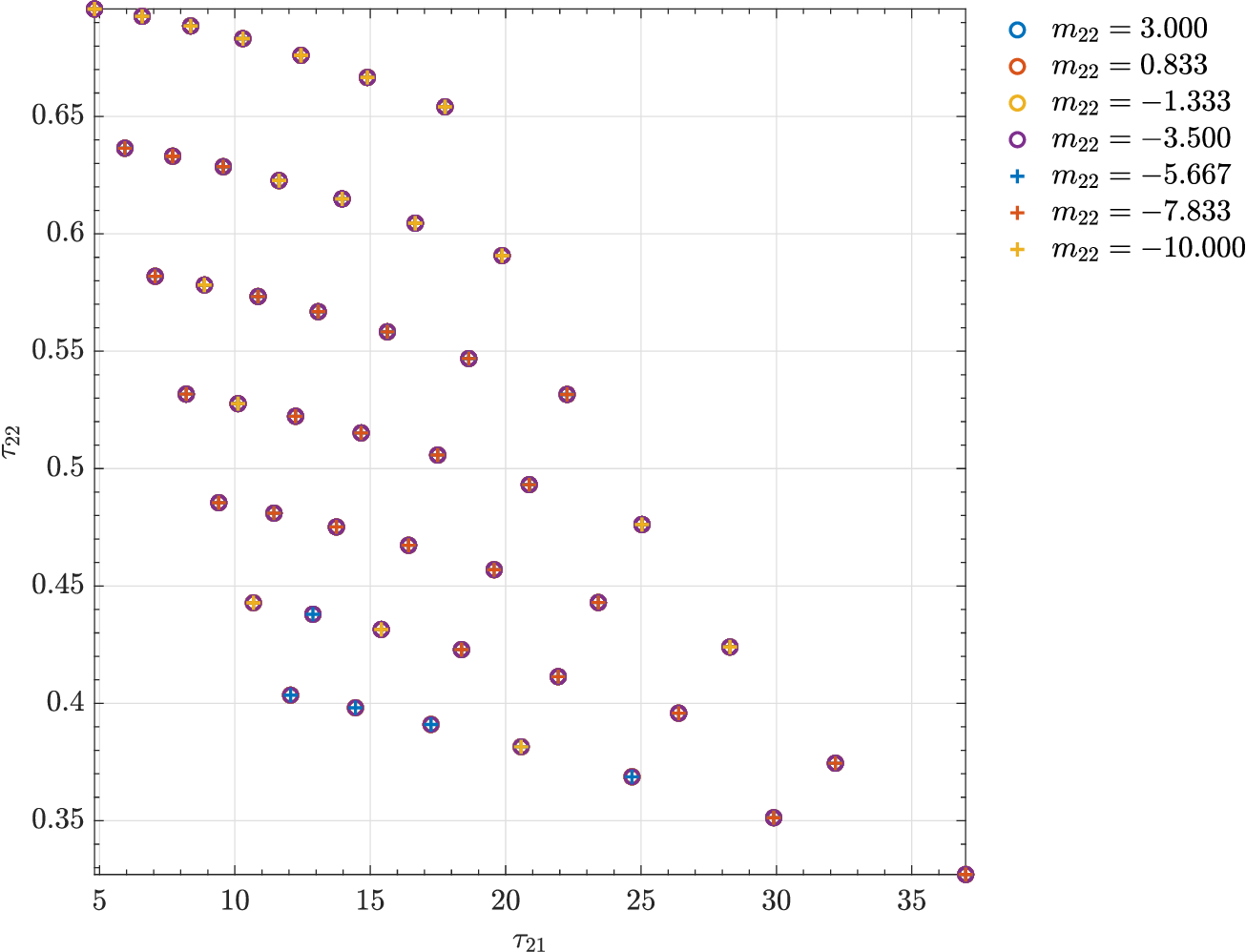}
\end{subfigure}
\caption{Values for \textbf{Case 3} of suitable solution pairs $\tau_{21}$, $\tau_{22}$ for combinations of bound pairs $m_{11}$, $m_{12}$ and $m_{21}$, $m_{22}$ (cont.)}\label{fig05bb}
\end{figure}

\end{document}